\pdfoutput=1

\documentclass[12pt,a4paper]{article}

\usepackage{ifthen}
\newboolean{pdflatex}
\setboolean{pdflatex}{true} 

\newboolean{articletitles}
\setboolean{articletitles}{true} 

\newboolean{uprightparticles}
\setboolean{uprightparticles}{false} 

\newboolean{inbibliography}
\setboolean{inbibliography}{false} 

\usepackage{microtype}
\usepackage{lineno}  
\usepackage{xspace} 

\usepackage{graphicx}  
\usepackage{color}
\usepackage{colortbl}
\graphicspath{{./figs/}} 

\usepackage{amsmath} 
\usepackage{amssymb}
\usepackage{amsfonts}
\usepackage{upgreek} 

\newcommand*\patchAmsMathEnvironmentForLineno[1]{%
\expandafter\let\csname old#1\expandafter\endcsname\csname #1\endcsname
\expandafter\let\csname oldend#1\expandafter\endcsname\csname
end#1\endcsname
 \renewenvironment{#1}%
   {\linenomath\csname old#1\endcsname}%
   {\csname oldend#1\endcsname\endlinenomath}%
}
\newcommand*\patchBothAmsMathEnvironmentsForLineno[1]{%
  \patchAmsMathEnvironmentForLineno{#1}%
  \patchAmsMathEnvironmentForLineno{#1*}%
}
\AtBeginDocument{%
\patchBothAmsMathEnvironmentsForLineno{equation}%
\patchBothAmsMathEnvironmentsForLineno{align}%
\patchBothAmsMathEnvironmentsForLineno{flalign}%
\patchBothAmsMathEnvironmentsForLineno{alignat}%
\patchBothAmsMathEnvironmentsForLineno{gather}%
\patchBothAmsMathEnvironmentsForLineno{multline}%
}

\usepackage{hyperref}    
\usepackage[all]{hypcap} 

\def\lhcb {\mbox{LHCb}\xspace}

\def\babar  {\mbox{BaBar}\xspace}
\def\belle  {\mbox{Belle}\xspace}

\ifthenelse{\boolean{uprightparticles}}%
{

 \def\Ppi         {\ensuremath{\uppi}\xspace}                 
                  
 \def\Prho        {\ensuremath{\uprho}\xspace}

 \def\Pomega      {\ensuremath{\upomega}\xspace}                 

 \def\PDelta      {\ensuremath{\Delta}\xspace}                 
 \def\PXi      {\ensuremath{\Xi}\xspace}                 
 \def\PLambda      {\ensuremath{\Lambda}\xspace}                 
 \def\PSigma      {\ensuremath{\Sigma}\xspace}                 
 \def\POmega      {\ensuremath{\Omega}\xspace}                 
 \def\PUpsilon      {\ensuremath{\Upsilon}\xspace}                 
 
 \def\PB      {\ensuremath{\mathrm{B}}\xspace}                 
                  
 \def\PD      {\ensuremath{\mathrm{D}}\xspace}

 \def\PK      {\ensuremath{\mathrm{K}}\xspace}

 \def\Pb      {\ensuremath{\mathrm{b}}\xspace}                 
 \def\Pc      {\ensuremath{\mathrm{c}}\xspace}                 
 \def\Pd      {\ensuremath{\mathrm{d}}\xspace}

 \def\Pi      {\ensuremath{\mathrm{i}}\xspace}

 \def\Pu      {\ensuremath{\mathrm{u}}\xspace}

}
{

 \def\Ppi         {\ensuremath{\pi}\xspace}                 
                  
 \def\Prho        {\ensuremath{\rho}\xspace}

 \def\Pomega      {\ensuremath{\omega}\xspace}                 
 \mathchardef\PDelta="7101
 \mathchardef\PXi="7104
 \mathchardef\PLambda="7103
 \mathchardef\PSigma="7106
 \mathchardef\POmega="710A
 \mathchardef\PUpsilon="7107
                  
 \def\PB      {\ensuremath{B}\xspace}                 
                  
 \def\PD      {\ensuremath{D}\xspace}

 \def\PK      {\ensuremath{K}\xspace}

 \def\Pb      {\ensuremath{b}\xspace}                 
 \def\Pc      {\ensuremath{c}\xspace}                 
 \def\Pd      {\ensuremath{d}\xspace}

 \def\Pi      {\ensuremath{i}\xspace}

 \def\Pu      {\ensuremath{u}\xspace}

}

\makeatletter
\ifcase \@ptsize \relax
  \newcommand{\miniscule}{\@setfontsize\miniscule{4}{5}}
\or
  \newcommand{\miniscule}{\@setfontsize\miniscule{5}{6}}
\or
  \newcommand{\miniscule}{\@setfontsize\miniscule{5}{6}}
\fi
\makeatother

\DeclareRobustCommand{\optbar}[1]{\shortstack{{\miniscule (\rule[.5ex]{1.25em}{.18mm})}
  \\ [-.7ex] $#1$}}

\def\uquark    {{\ensuremath{\Pu}}\xspace}

\def\dquark    {{\ensuremath{\Pd}}\xspace}

\def\cquark    {{\ensuremath{\Pc}}\xspace}

\def\bquark    {{\ensuremath{\Pb}}\xspace}

\def\pion   {{\ensuremath{\Ppi}}\xspace}

\def\pip    {{\ensuremath{\pion^+}}\xspace}
\def\pim    {{\ensuremath{\pion^-}}\xspace}
\def\pipm   {{\ensuremath{\pion^\pm}}\xspace}
\def\pimp   {{\ensuremath{\pion^\mp}}\xspace}

\def\kaon    {{\ensuremath{\PK}}\xspace}
  \def\Kbar    {{\kern 0.2em\overline{\kern -0.2em \PK}{}}\xspace}

\def\KorKbar    {\kern 0.18em\optbar{\kern -0.18em K}{}\xspace}

\def\Kp      {{\ensuremath{\kaon^+}}\xspace}
\def\Km      {{\ensuremath{\kaon^-}}\xspace}
\def\Kpm     {{\ensuremath{\kaon^\pm}}\xspace}
\def\Kmp     {{\ensuremath{\kaon^\mp}}\xspace}
\def\KS      {{\ensuremath{\kaon^0_{\rm\scriptscriptstyle S}}}\xspace}

\def\Kstarz  {{\ensuremath{\kaon^{*0}}}\xspace}
\def\Kstarzb {{\ensuremath{\Kbar{}^{*0}}}\xspace}
\def\Kstar   {{\ensuremath{\kaon^*}}\xspace}

\def\Kstarpm {{\ensuremath{\kaon^{*\pm}}}\xspace}

  \def\Dbar    {{\kern 0.2em\overline{\kern -0.2em \PD}{}}\xspace}
\def\D       {{\ensuremath{\PD}}\xspace}

\def\DorDbar    {\kern 0.18em\optbar{\kern -0.18em D}{}\xspace}
\def\Dz      {{\ensuremath{\D^0}}\xspace}
\def\Dzb     {{\ensuremath{\Dbar{}^0}}\xspace}

\def\Dstar   {{\ensuremath{\D^*}}\xspace}

\def\Dstarz  {{\ensuremath{\D^{*0}}}\xspace}
\def\Dstarzb {{\ensuremath{\Dbar{}^{*0}}}\xspace}

\def\Dstarpm {{\ensuremath{\D^{*\pm}}}\xspace}
\def\Dstarmp {{\ensuremath{\D^{*\mp}}}\xspace}

\def\B       {{\ensuremath{\PB}}\xspace}
\def\Bbar    {{\ensuremath{\kern 0.18em\overline{\kern -0.18em \PB}{}}}\xspace}

\def\BorBbar    {\kern 0.18em\optbar{\kern -0.18em B}{}\xspace}

\def\Bu      {{\ensuremath{\B^+}}\xspace}
\def\Bub     {{\ensuremath{\B^-}}\xspace}
\def\Bp      {{\ensuremath{\Bu}}\xspace}
\def\Bm      {{\ensuremath{\Bub}}\xspace}
\def\Bpm     {{\ensuremath{\B^\pm}}\xspace}

  \def\Y#1S{\ensuremath{\PUpsilon{(#1S)}}\xspace}

\def\Lbar        {{\ensuremath{\kern 0.1em\overline{\kern -0.1em\PLambda}}}\xspace}
\def\LorLbar    {\kern 0.18em\optbar{\kern -0.18em \PLambda}{}\xspace}

\def\to                 {\ensuremath{\rightarrow}\xspace}

\def\CP                {{\ensuremath{C\!P}}\xspace}

\def\Vud  {{\ensuremath{V_{\uquark\dquark}}}\xspace}
\def\Vcd  {{\ensuremath{V_{\cquark\dquark}}}\xspace}

\def\Vub  {{\ensuremath{V_{\uquark\bquark}}}\xspace}
\def\Vcb  {{\ensuremath{V_{\cquark\bquark}}}\xspace}

\def\AT#1     {\ensuremath{A_{\mathrm{T}}^{#1}}\xspace}           

\def\C#1      {\ensuremath{\mathcal{C}_{#1}}\xspace}                       
\def\Cp#1     {\ensuremath{\mathcal{C}_{#1}^{'}}\xspace}                    
\def\Ceff#1   {\ensuremath{\mathcal{C}_{#1}^{\mathrm{(eff)}}}\xspace}        
\def\Cpeff#1  {\ensuremath{\mathcal{C}_{#1}^{'\mathrm{(eff)}}}\xspace}       
\def\Ope#1    {\ensuremath{\mathcal{O}_{#1}}\xspace}                       
\def\Opep#1   {\ensuremath{\mathcal{O}_{#1}^{'}}\xspace}                    

\newcommand{\tev}{\ifthenelse{\boolean{inbibliography}}{\ensuremath{~T\kern -0.05em eV}\xspace}{\ensuremath{\mathrm{\,Te\kern -0.1em V}}}\xspace}
\newcommand{\gev}{\ensuremath{\mathrm{\,Ge\kern -0.1em V}}\xspace}
\newcommand{\mev}{\ensuremath{\mathrm{\,Me\kern -0.1em V}}\xspace}
\newcommand{\kev}{\ensuremath{\mathrm{\,ke\kern -0.1em V}}\xspace}
\newcommand{\ev}{\ensuremath{\mathrm{\,e\kern -0.1em V}}\xspace}
\newcommand{\gevc}{\ensuremath{{\mathrm{\,Ge\kern -0.1em V\!/}c}}\xspace}
\newcommand{\mevc}{\ensuremath{{\mathrm{\,Me\kern -0.1em V\!/}c}}\xspace}
\newcommand{\gevcc}{\ensuremath{{\mathrm{\,Ge\kern -0.1em V\!/}c^2}}\xspace}
\newcommand{\gevgevcccc}{\ensuremath{{\mathrm{\,Ge\kern -0.1em V^2\!/}c^4}}\xspace}
\newcommand{\mevcc}{\ensuremath{{\mathrm{\,Me\kern -0.1em V\!/}c^2}}\xspace}

\def\mum  {\ensuremath{{\,\upmu\rm m}}\xspace}

\def\invfb   {\ensuremath{\mbox{\,fb}^{-1}}\xspace}

\newcommand{\chisq}{\ensuremath{\chi^2}\xspace}

\newcommand{\chisqip}{\ensuremath{\chi^2_{\rm IP}}\xspace}

\def\gsim{{~\raise.15em\hbox{$>$}\kern-.85em
          \lower.35em\hbox{$\sim$}~}\xspace}
\def\lsim{{~\raise.15em\hbox{$<$}\kern-.85em
          \lower.35em\hbox{$\sim$}~}\xspace}

\def\ptot       {\mbox{$p$}\xspace}
\def\pt         {\mbox{$p_{\rm T}$}\xspace}

\def\evtgen     {\mbox{\textsc{EvtGen}}\xspace}

\def\geant      {\mbox{\textsc{Geant4}}\xspace}

\def\photos     {\mbox{\textsc{Photos}}\xspace}

\def\pythia     {\mbox{\textsc{Pythia}}\xspace}

\def\tell1  {TELL1\xspace}
\def\ukl1   {UKL1\xspace}

\usepackage{cite} 
\usepackage{mciteplus}
\usepackage{multirow}

\def\BptoDKp          {\mbox{$\Bp \rightarrow \D \Kp$}}
\def\BmtoDKm          {\mbox{$\Bm \rightarrow \D \Km$}}

\def\BpmtoDKpm        {\mbox{$\Bpm \rightarrow \D \Kpm$}}

\def\BpmtoDpipm       {\mbox{$\Bpm \rightarrow \D \pipm$}}
\def\BpmtoDKspipiKpm  {\mbox{$\Bpm \rightarrow \D (\rightarrow \KS \pip \pim ) \Kpm$}}
\def\BpmtoDKspipipipm {\mbox{$\Bpm \rightarrow \D (\rightarrow \KS \pip \pim ) \pipm$}}

\def\DtoKspipi        {\mbox{$\D \rightarrow \KS \pip \pim$}}
\def\DztoKspipi       {\mbox{$\Dz \rightarrow \KS \pip \pim$}}
\def\DzbtoKspipi      {\mbox{$\Dzb \rightarrow \KS \pip \pim$}}

\def\gam	      {$\gamma$}

\newcommand{\plot}[2][width=.95\textwidth]{
  \ifpdf
  \IfFileExists{\finkbase/png/#2.png}
  {
    \includegraphics[#1,type=png,ext=.png,read=.png]{\finkbase/png/#2}
  }
  {
    \IfFileExists{\finkbase/jpg/#2.jpg}
    {
      \includegraphics[#1,type=jpg,ext=.jpg,read=.jpg]{\finkbase/jpg/#2}
    }
    {
      \includegraphics[#1,type=png,ext=.png,read=.png]{format/blank}
    }
  }
  \else
  \IfFileExists{\finkbase/eps/#2.eps}{
    \includegraphics[#1,type=eps,ext=.eps,read=.eps]{\finkbase/eps/#2}
  }{
    \includegraphics[#1,type=eps,ext=.eps,read=.eps]{format/blank}
  }
  \fi
}

\definecolor{orange}{rgb}{1,0.5,0}

\newcommand{\re}[2][()] {\ifthenelse{\equal{#1}{()}}{{\ensuremath{{\rm \, Re}}\left(#2\right)}}
                                                    {{\ensuremath{{\rm \, Re}}\left[#2\right]}}}
\newcommand{\im}[2][()] {\ifthenelse{\equal{#1}{()}}{{\ensuremath{{\rm \, Im}}\left(#2\right)}}
                                                    {{\ensuremath{{\rm \, Im}}\left[#2\right]}}}

\newcommand{\GeV}       {\ensuremath{\rm \, GeV}}

\newcommand{\mrow}[2]   {\multirow{#1}{*}{#2}}
\newcommand{\mcol}[2]   {\multicolumn{#1}{l}{#2}}

\newcommand{\xypm}      {$(x_{\pm},y_{\pm})$\xspace}

\newcommand{\TMT}       {(\times 10^{-3})}

\newcommand{\mygauss}[3]  {
                        \ifthenelse{\isempty{#3}}
                        {G\left( #1; \mu^{#1,#2}     , \sigma^{#1,#2}      \right)}
                        {G\left( #1; \mu^{#1,#2}_{#3}, \sigma^{#1,#2}_{#3} \right)}
                        }
\newcommand{\bgauss}[3] {
                        \ifthenelse{\isempty{#3}}
                        {G_b\left( #1; \mu^{#1,#2}     , \sigma_l^{#1,#2}     , \sigma_r^{#1,#2}      \right)}
                        {G_b\left( #1; \mu^{#1,#2}_{#3}, \sigma_l^{#1,#2}_{#3}, \sigma_r^{#1,#2}_{#3} \right)}
                        }
\newcommand{\gaussC}[4] {
                        \ifthenelse{\isempty{#4}}
                        {G_c\left( #1,#2;\mu^{#1,#3}     ,\sigma^{#1,#3}     ,
                        \mu^{#2,#3}     ,\sigma^{#2,#3}     ,\kappa     \right)}
                        {G_c\left( #1,#2;\mu^{#1,#3}_{#4},\sigma^{#1,#3}_{#4},
                        \mu^{#2,#3}_{#4},\sigma^{#2,#3}_{#4},\kappa_{#4}\right)}
                        }
\newcommand{\johnson}[3]{
                        \ifthenelse{\isempty{#3}}
                        {J_{S_U}\left(#1;\mu^{#1,#2}     ,\sigma^{#1,#2}     ,
                        \gamma^{#1,#2}     ,\delta^{#1,#2}     \right)}
                        {J_{S_U}\left(#1;\mu^{#1,#2}_{#3},\sigma^{#1,#2}_{#3},
                        \gamma^{#1,#2}_{#3},\delta^{#1,#2}_{#3}\right)}
                        }
\newcommand{\argus}[3]  {
                        \ifthenelse{\isempty{#3}}
                        {B\left(#1;m_\pi,\xi^{#1,#2}     \right)}
                        {B\left(#1;m_\pi,\xi^{#1,#2}_{#3}\right)}
                        }
\newcommand{\mitj}[1]   {
                        \ifthenelse{\isempty{#1}}
                        {\langle \mu      \rangle}
                        {\langle \mu^{#1} \rangle}
                        }
\newcommand{\desv}[1]   {
                        \ifthenelse{\isempty{#1}}
                        {\langle \sigma      \rangle}
                        {\langle \sigma^{#1} \rangle}
                        }
\newcommand{\kst}[3]    {
                        \ifthenelse{\isempty{#2}}
                        {$K^{\star #1}     (#3)$\xspace}
                        {$K^{\star #1}_{#2}(#3)$\xspace}
                        }

\begin{document}

\renewcommand{\thefootnote}{\fnsymbol{footnote}}
\setcounter{footnote}{1}

\begin{titlepage}
\pagenumbering{roman}

\vspace*{-1.5cm}
\centerline{\large EUROPEAN ORGANIZATION FOR NUCLEAR RESEARCH (CERN)}
\vspace*{1.5cm}
\hspace*{-0.5cm}
\begin{tabular*}{\linewidth}{lc@{\extracolsep{\fill}}r}
\ifthenelse{\boolean{pdflatex}}
{\vspace*{-2.7cm}\mbox{\!\!\!\includegraphics[width=.14\textwidth]{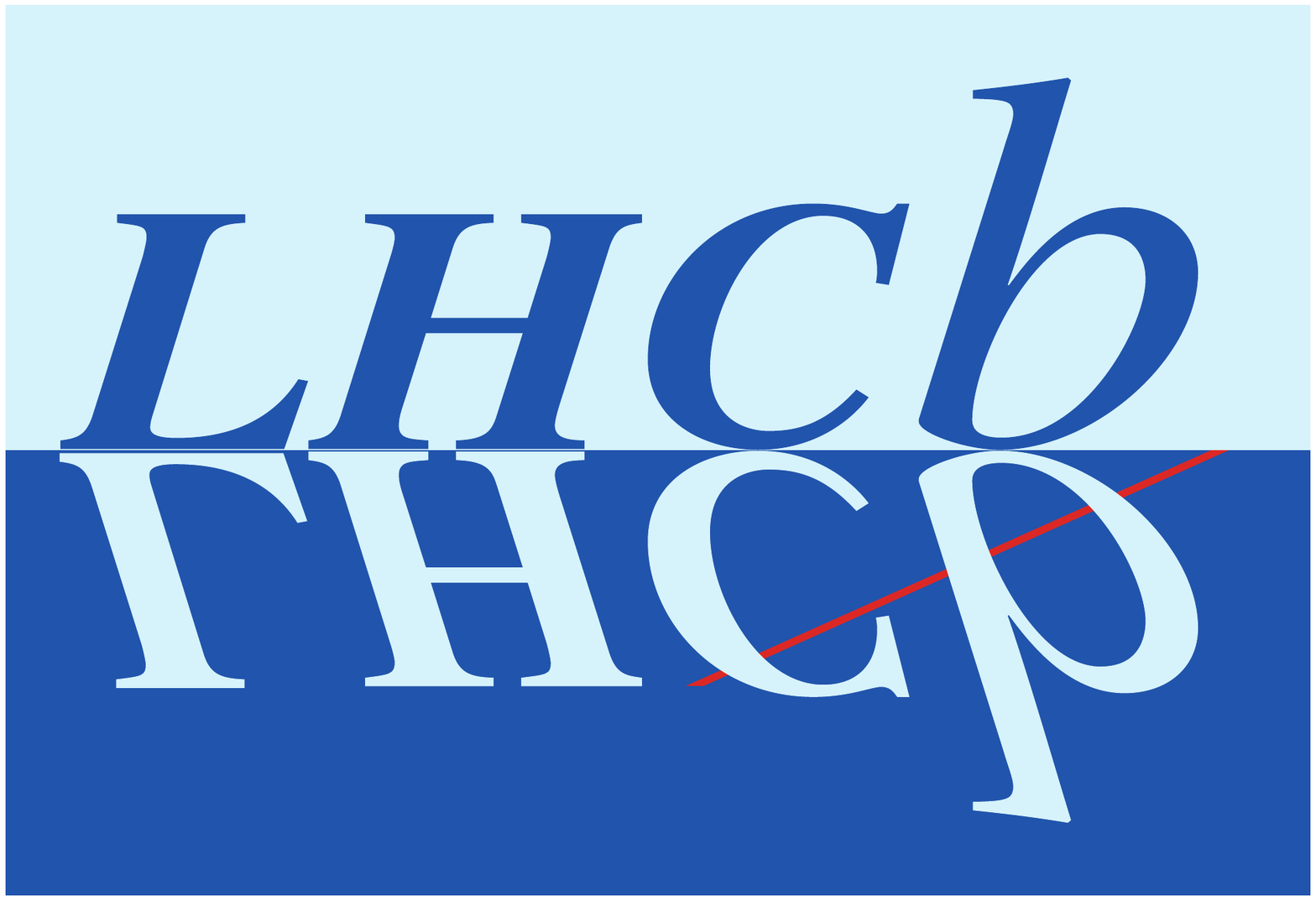}} & &}
{\vspace*{-1.2cm}\mbox{\!\!\!\includegraphics[width=.12\textwidth]{lhcb-logo.eps}} & &}
\\
 & & CERN-PH-EP-2014-166 \\  
 & & LHCb-PAPER-2014-017 \\  
 & & July 22, 2014 \\ 
 & & \\

\end{tabular*}

\vspace*{1.cm}

{\bf\boldmath\huge
\begin{center}
Measurement of \CP violation and constraints on the CKM angle \gam\ in $\Bpm\rightarrow\D \Kpm$ with $D \rightarrow \KS \pip \pim$ decays
\end{center}
}

\vspace*{0.5cm}

\begin{center}
The LHCb collaboration\footnote{Authors are listed at the end of this paper.}
\end{center}

\vspace{\fill}

\begin{abstract}
  \noindent
A model-dependent amplitude analysis of $\Bpm\rightarrow\D \Kpm$ with $D \rightarrow \KS \pip \pim$ decays is performed
using proton-proton collision data, corresponding to an integrated luminosity of $1\invfb$, 
recorded by \lhcb at a centre-of-mass energy of $7\tev$ in $2011$. 
Values of the \CP violation observables $x_{\pm}$ and $y_{\pm}$, which are sensitive to the CKM angle \gam, are measured to be
\begin{align}
x_- &= +0.027 \pm 0.044 ~^{+0.010}_{-0.008} \pm 0.001, \notag \\
y_- &= +0.013 \pm 0.048 ~^{+0.009}_{-0.007} \pm 0.003, \notag \\
x_+ &= -0.084 \pm 0.045 \pm 0.009          \pm 0.005, \notag \\
y_+ &= -0.032 \pm 0.048 ~^{+0.010}_{-0.009} \pm 0.008, \notag
\end{align}
where the first uncertainty is statistical, the second systematic and the third arises from the uncertainty 
of the $\D \rightarrow \KS \pip \pim$ amplitude model.  
The value of \gam\ is determined to be $(84^{+49}_{-42})^\circ$, including all sources of uncertainty.
Neutral \D meson mixing is found to have negligible effect.

\end{abstract}

\vspace*{0.5cm}

\begin{center}
  Published in Nucl.~Phys.~B
\end{center}

\vspace{\fill}

{\footnotesize
\centerline{\copyright~CERN on behalf of the \lhcb collaboration, license \href{http://creativecommons.org/licenses/by/4.0/}{CC-BY-4.0}.}}
\vspace*{2mm}

\end{titlepage}

\newpage
\setcounter{page}{2}
\mbox{~}

\cleardoublepage

\renewcommand{\thefootnote}{\arabic{footnote}}
\setcounter{footnote}{0}


\pagestyle{plain} 
\setcounter{page}{1}
\pagenumbering{arabic}

\section{Introduction}
\label{sec:Introduction}
\newcommand{\bl}{\hphantom{-}}
\newcommand{\bn}{\hphantom{9}}
The CKM phase $\gamma$
($\gamma \equiv \mathrm{arg} \left[- \Vud \Vub^* / \Vcd \Vcb^*  \right]$, also known as $\phi_3$)
is the angle of the CKM unitarity triangle that is least constrained by direct measurements.
The precise determination of $\gamma$ is an important aim of current flavour physics experiments.
It can be measured directly in tree-level processes, for example in \BpmtoDKpm\ decays
where \D\ is a superposition of the flavour eigenstates \Dz\ and \Dzb decaying into the same final
state. Sensitivity to \gam\ arises from the interference between $\bquark \rightarrow \uquark$ and
$\bquark \rightarrow \cquark$ quark transitions.
Since \BpmtoDKpm\ decays are expected to be insensitive to physics processes beyond the Standard Model (SM),
this measurement provides a reference value against which other observables,
potentially affected by physics beyond the SM, can be compared.

The determination of \gam\ (using \BpmtoDKpm\ decays) from an amplitude analysis of the \D\ meson decay to the three-body quasi-self-conjugate
$\KS \pip \pim$ final state
was first proposed in Refs.~\cite{Giri:2003ty, BondarGGSZ}.  The method requires knowledge of
the \DtoKspipi\ decay amplitude
across the phase space and, in particular, its strong phase variation.
The model-dependent approach, as used in 
Refs.~\cite{Aubert:2005iz, Aubert:2008bd, delAmoSanchez:2010rq, Poluektov:2004mf, Poluektov:2006ia, Poluektov:2010wz}, 
implements a model to describe the \D\ decay amplitude over the phase space.  
This unbinned method allows for full exploitation of the statistical power of the data.
A model-independent strategy, employed by the \lhcb~\cite{LHCb-PAPER-2012-027}
and \belle~\cite{Aihara:2012aw} collaborations,
uses CLEO measurements~\cite{Libby:2010nu} of the \D\ decay strong phase difference in bins across the
phase space.

Neglecting the effects of charm mixing,
the amplitude for \BpmtoDKspipiKpm\ decays can be written as a superposition of Cabibbo favoured and suppressed contributions,
\begin{align}
\mathcal{A}_{\Bm} &\sim     {A}_f +   r_{\B} e^{i(\delta_{\B} - \gamma)} \bar{A}_f, \label{eq.intro.Ap} \\
\mathcal{A}_{\Bp} &\sim \bar{A}_f +   r_{\B} e^{i(\delta_{\B} + \gamma)}     {A}_f, \notag 
\end{align}
where $r_{\B}$ is the magnitude of the ratio of the interfering \Bpm\ decay amplitudes, $\delta_{\B}$ is the strong phase
difference between them, and $\gamma$ is the \CP-violating weak phase.
The amplitudes of the \Dz\ and \Dzb\ mesons decaying into the common final
state $f$, \mbox{$A_f \equiv \left\langle f \vphantom{\bar{D}^0} \right| \mathcal{H} \left| \vphantom{\bar{D}^0} \Dz \right\rangle$} and
\mbox{$\bar{A}_f \equiv \left\langle f \vphantom{\bar{D}^0} \right| \mathcal{H} \left| \Dzb \right\rangle$}, respectively,
depend on two squared invariant masses of pairs of the three final state
particles, chosen to be
\mbox{$m_{+}^2 \equiv m_{\KS \pip}^2$} and \mbox{$m_{-}^2 \equiv m_{\KS \pim}^2$}.
Assuming that no direct \CP violation exists in the $D$ meson decay, the amplitudes $A_f$ and $\bar{A}_f$ are
related by $\bar{A}_f( m^2_+, m^2_- ) = A_f( m^2_-, m^2_+ )$.
A direct determination of $r_{\B}$, $\delta_{\B}$ and \gam\ can lead to bias~\cite{Aubert:2005iz}, 
and hence the Cartesian \CP violation observables, \mbox{$x_{\pm} = r_{\B} \cos{(\delta_{\B} \pm \gamma)}$} and
\mbox{$y_{\pm} = r_{\B} \sin{(\delta_{\B} \pm \gamma)}$}, are used, 
where the ``$+$'' and ``$-$'' indices correspond to \Bp\ and \Bm\ decays, respectively.

This paper reports measurements of \xypm made
using \BpmtoDKspipiKpm\ decays selected from $pp$ collision data, corresponding to an integrated luminosity of
$1\invfb$, recorded by \lhcb at a centre-of-mass energy of $7\tev$ in $2011$.
The data set is identical to that used in Ref.~\cite{LHCb-PAPER-2012-027}.
The measured values of \xypm place constraints on the CKM angle \gam.

\section{The \lhcb detector}
\label{sec:Detector}
The \lhcb detector~\cite{Alves:2008zz} is a single-arm forward
spectrometer covering the \mbox{pseudorapidity} range $2<\eta <5$,
designed for the study of particles containing \bquark or \cquark
quarks. The detector includes a high-precision tracking system
consisting of a silicon-strip vertex detector surrounding the $pp$
interaction region, a large-area silicon-strip detector located
upstream of a dipole magnet with a bending power of about
$4{\rm\,Tm}$, and three stations of silicon-strip detectors and straw
drift tubes placed downstream of the magnet.
The combined tracking system provides a momentum measurement with
a relative uncertainty that varies from 0.4\% at low momentum, \ptot, to 0.6\% at 100\gevc,
and an impact parameter measurement with a resolution of 20\mum for
charged particles with large transverse momentum, \pt. Different types of charged hadrons are distinguished using information
from two ring-imaging Cherenkov detectors~\cite{LHCb-DP-2012-003}, providing particle
identification (PID) information. 
Photon, electron and
hadron candidates are identified by a calorimeter system consisting of
scintillating-pad and preshower detectors, an electromagnetic
calorimeter and a hadronic calorimeter. Muons are identified by a
system composed of alternating layers of iron and multiwire
proportional chambers.

The trigger consists of a hardware stage, based on information from the calorimeter
and muon systems, followed by a software stage, which applies a full event
reconstruction.  
The software trigger requires a two-, three- or four-track
secondary vertex with a large sum \pt of
the tracks and a significant displacement from any primary $pp$ interaction vertex~(PV).
At least one track should also have large \pt and \chisqip with respect to any primary interaction,
where \chisqip is defined as the difference in \chisq of a given PV reconstructed
with and without the considered track. A multivariate algorithm~\cite{BBDT} is used to identify secondary vertices consistent with decays of \bquark hadrons.

Large samples of simulated \BpmtoDKspipiKpm\ and \BpmtoDKspipipipm\ decays are used in this study,
along with simulated samples of various background decays.
In the simulation, $pp$ collisions are generated using
\pythia~6.4~\cite{Sjostrand:2006za} with a specific \lhcb
configuration~\cite{LHCb-PROC-2010-056}.  Decays of hadronic particles
are described by \evtgen~\cite{Lange:2001uf}, in which final state
radiation is generated using \photos~\cite{Golonka:2005pn}. The
interaction of the generated particles with the detector and its
response are implemented using the \geant toolkit~\cite{Allison:2006ve, *Agostinelli:2002hh}
as described in Ref.~\cite{LHCb-PROC-2011-006}.

\section{Candidate selection and sources of background}
\label{sec:Selection}
The criteria used to select \BpmtoDKspipiKpm\ and \BpmtoDKspipipipm\ candidate decays from the data 
are described below.
The \BpmtoDKspipipipm\ decays are used to measure the acceptance over phase space, 
as they have almost identical topologies to \BpmtoDKspipiKpm\ decays, but a much higher branching fraction~\cite{PDG2012}.
Apart from the \Bpm candidate invariant mass range, the selection requirements are identical to those used in Ref.~\cite{LHCb-PAPER-2012-027} and are summarised here for completeness.

Candidate \KS mesons are reconstructed from two oppositely charged well-measured tracks; those 
with tracks reconstructed in the silicon vertex detector are known as {\it long} candidates and those 
with tracks that cannot be formed in the vertex detector are known 
as {\it downstream} candidates.
A requirement of \chisqip greater than 16 (4) with respect to the PV is made for the {\it long} ({\it downstream}) pion tracks.
The PV of each candidate \Bpm meson decay is chosen to be the one yielding the minimum \chisqip.
To reduce background from random track combinations, 
the cosine of the angle between the momentum direction of the \KS meson candidate and the direction vector from the PV to its decay vertex is required to be greater than 0.99.

The \KS candidates are combined with two oppositely charged tracks to reconstruct \D meson candidates; 
the tracks combined with a {\it long} ({\it downstream}) candidate must have \chisqip greater than 9 (16) with respect to the PV. 
For all \D meson candidates, requirements of \chisqip greater than 9 with respect to the PV and 
cosine of the angle between the momentum and direction vectors greater than 0.99
are made.
It is required that the vertex separation \chisq between the reconstructed 
\D and \KS meson decay vertices is greater than 100,
where the vertex separation \chisq is defined as the change in \chisq of a vertex which is reconstructed including 
the particles originally contributing to the other vertex.
The reconstructed \D\ meson candidate invariant mass is required to be within $\pm 25\mevcc$
around the known value~\cite{PDG2012}.
The \KS candidate invariant mass must be within $\pm 15\mevcc$
around the known value~\cite{PDG2012} after a refit to constrain the \D meson mass~\cite{Hulsbergen:2005pu}.

The \Bpm meson candidates are reconstructed from the combination of a \D meson candidate with a pion or kaon 
directly from the \Bpm vertex, hereafter called the ``bachelor'' track.
The bachelor track is required to have \chisqip greater than 25 with respect to the PV.
To separate \BpmtoDKpm\ and \BpmtoDpipm\ decays, 
good discrimination between pions and kaons is required using PID information.
The \chisqip of the reconstructed \Bpm candidate with respect to the PV is required to be less than 9,
and for {\it long} ({\it downstream}) candidates the cosine of the angle between 
its momentum and direction vectors must be greater than 0.9999 (0.99995).
The \Bpm vertex separation \chisq with respect to the PV must be greater than 169.
In addition, the reconstructed \D meson decay vertex is required to have a larger longitudinal displacement from the PV than the \Bpm decay vertex.

Each selected candidate decay is refitted with additional constraints on the \KS\ and \D\ meson masses and
on the pointing of the \B\ momentum to the PV, so that improved resolution in the phase space
of the \D\ decay is obtained.  A refit quality requirement of \chisq per degree of freedom less than 5 is made.
If more than one selected candidate is found to originate from the same $pp$ collision event,
 the candidate with the lowest value of refit \chisq per degree of freedom is retained.

Several sources of potential background are studied using simulation.
These include two categories of combinatorial background: 
a real \DtoKspipi\ decay combined with a random bachelor track (random $Dh$),
or a \DtoKspipi\ candidate reconstructed with at least one random final state track (combinatorial \D).
Cross-feed background arises from \BpmtoDKspipipipm\ decays misidentified as \BpmtoDKspipiKpm\ decays (or vice versa),
and contributes a large fraction of the selected \BpmtoDKspipiKpm\ candidates.
Partially reconstructed candidates from decay modes containing a \DtoKspipi\ decay,
such as \mbox{$\Bpm \rightarrow \Dstar h^{\pm}$}
(where \Dstar represents \Dstarz or \Dstarzb and $h^{\pm}$ represents a $\Kpm$ or $\pipm$),
\mbox{$\PB_{(s)} \rightarrow \D \Kstar$} (where $\PB_{(s)}$ represents $B^{0}_{(s)}$ or $\kern 0.18em\overline{\kern -0.18em \PB}{}^{0}_{(s)}$ and \Kstar represents \Kstarz or \Kstarzb) and
\mbox{$\Bpm \rightarrow \D \Prho^{\pm}$} decays,
are also expected to contribute.
The contributions from charmless \Bpm\ decays, \mbox{$\Bpm \rightarrow \D (\rightarrow \KS \Kpm \pimp ) h^{\pm}$} decays,
\mbox{$\Bpm \rightarrow \D (\rightarrow \KS \Kp \Km ) h^{\pm}$} decays and
\mbox{$\Bpm \rightarrow \D (\rightarrow \pip \pim h^+ h^- ) h^{\pm}$} decays
are found to be negligible.

\section{Analysis strategy}
\label{sec:Fit}
\begin{figure}[tb]
  \begin{center}
    \includegraphics[width=0.49\linewidth]{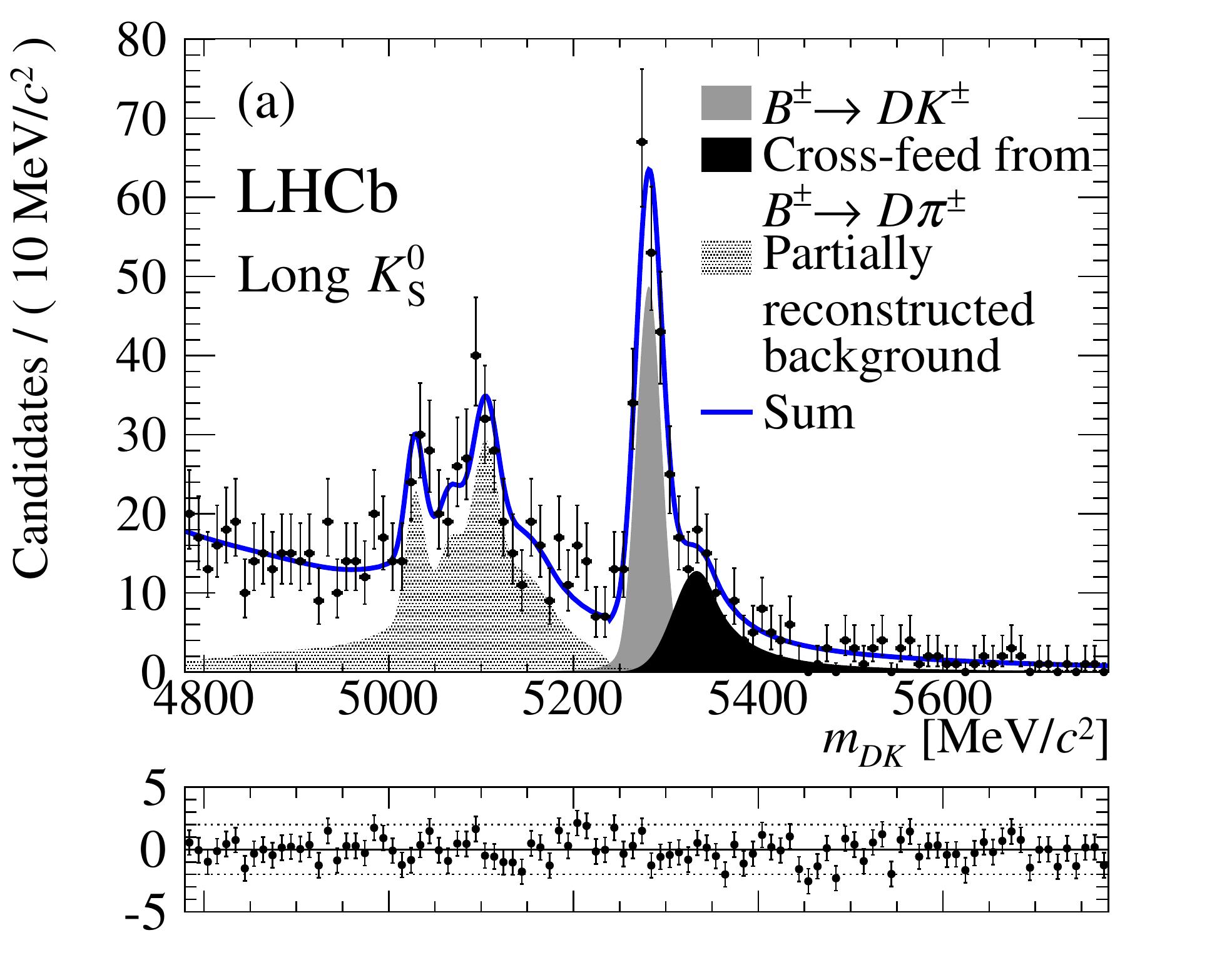}
    \includegraphics[width=0.49\linewidth]{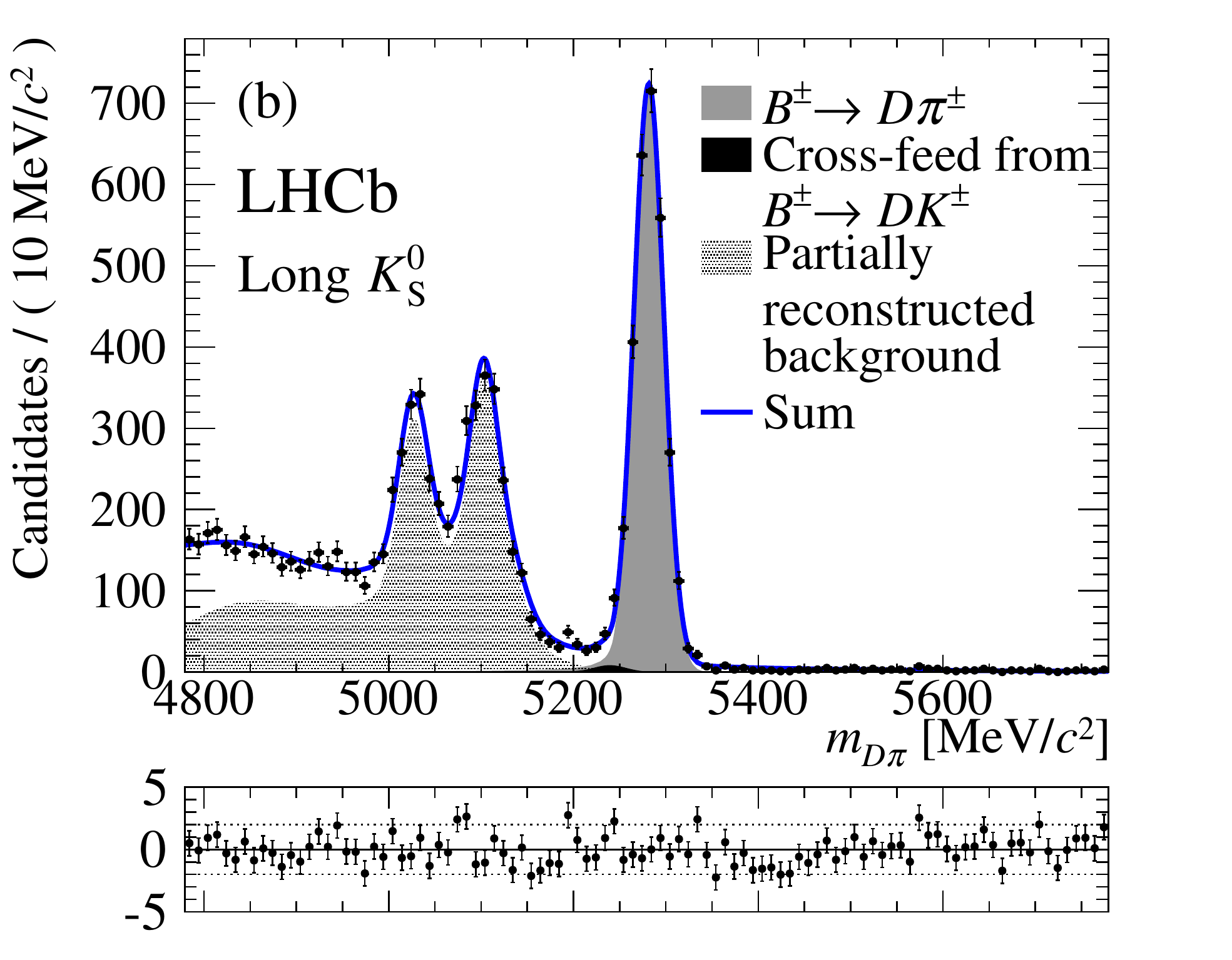}\\
    \includegraphics[width=0.49\linewidth]{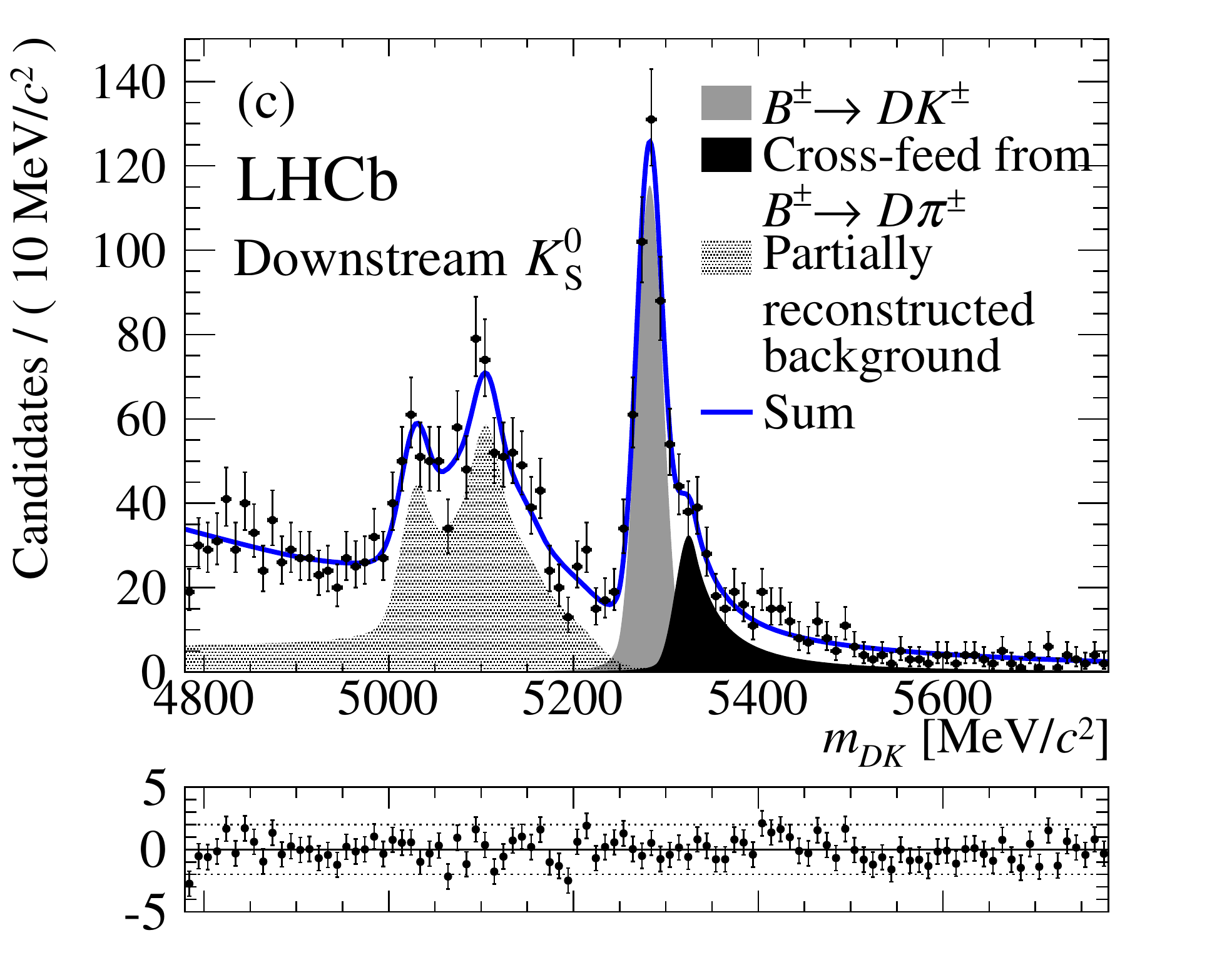}
    \includegraphics[width=0.49\linewidth]{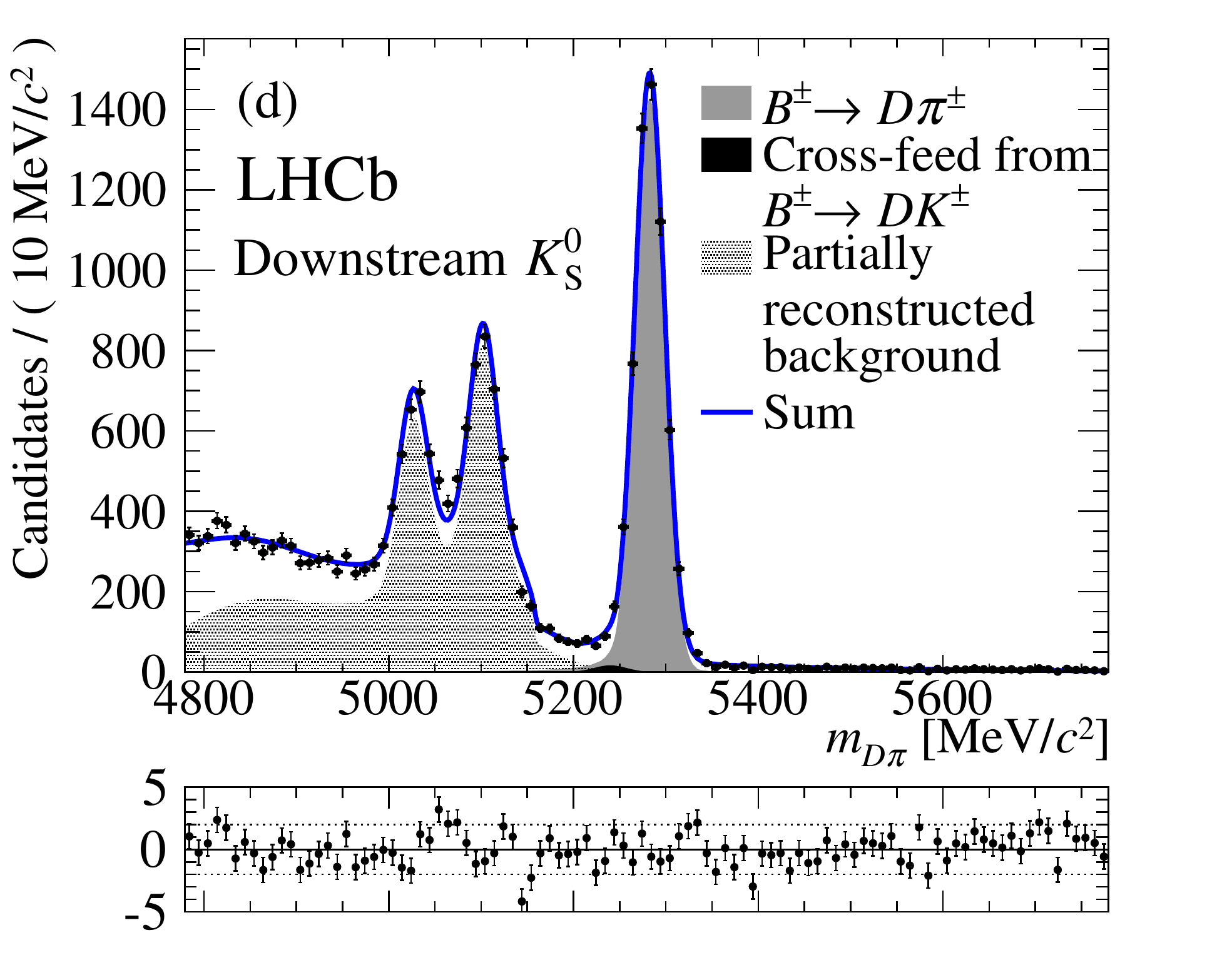}\\
    \vspace*{-0.5cm}
  \end{center}
  \caption{
    \small 
    Invariant mass distributions for (a) \BpmtoDKspipiKpm\ {\it long},  (b) \BpmtoDKspipipipm\ {\it long}, 
    (c) \BpmtoDKspipiKpm\ {\it downstream} and (d) \BpmtoDKspipipipm\ {\it downstream} candidates.
    The fit results, including signal and background components, are superimposed.
    The lower plots are normalised residual distributions.
    }
  \label{fig:massFits}
\end{figure}

The analysis is performed in two distinct parts.
The fractions of signal and background
are determined with a phase-space integrated fit to the invariant
mass distributions, $m_{Dh}$, of selected \BpmtoDKspipiKpm\ and \BpmtoDKspipipipm\ candidates, 
shown in Fig.~\ref{fig:massFits}.
This is followed by a fit to determine the \CP violation observables \xypm and the
variation in efficiency over the phase space of the \DtoKspipi\ decay.  The relative signal and background yields and the
parameters of the \Bpm\ invariant mass probability distribution functions (PDFs) are fixed to the values determined in the first stage.

\subsection{Invariant mass fit of \textbf{\textit{B}}$^\pm$ candidates}
\label{sec:MassFit}
An unbinned extended maximum likelihood fit to the invariant mass distributions of the \Bpm candidates
determines the signal and background fractions.
The samples of \BpmtoDKspipiKpm\ and \BpmtoDKspipipipm\ candidates are fitted simultaneously in
an invariant mass range of $4779\mevcc < m_{Dh} < 5779 \mevcc$.
The {\it long} and {\it downstream} candidates are fitted separately.

For the fit to the \BpmtoDKspipiKpm\ invariant mass distribution, the total PDF is composed of a signal and
several background components.  
The signal (\BpmtoDKpm) is described by the sum of a Crystal Ball~\cite{Skwarnicki:1986xj} and a Gaussian function with common means.
The Crystal Ball tail parameters, 
the width of the Gaussian function
and the relative fractions of both functions  
are fixed to values obtained from simulated data.  
An exponential function describes the two categories of combinatorial background candidates.  
Cross-feed candidates are characterised by a Crystal Ball function with tails on both upper and lower sides.
The mean and tail parameters of the function are fixed to results from simulation.  
Partially reconstructed background contributions are described by various functions with parameters fixed to values obtained from simulation.  
Both \PB and \Bpm decays that give rise to candidates with similar invariant mass distributions are described by a single fit component: 
the candidates from $\Bpm \rightarrow \Dstar \Kpm$ and $\PB \rightarrow \Dstarmp \Kpm$
decays are both described using the sum of two pairs of Gaussian functions,
where the Gaussian functions in each pair
have a common mean and independent widths.  
For the combined background contribution from partially reconstructed $\Bpm \rightarrow \Dstar \pipm$ and
$\PB \rightarrow \Dstarmp \pipm$ decays, labelled $\Dstar \pi$, the sum of two Crystal Ball functions, each with tails on both upper and lower sides, is used.  
A background composed of candidates from $\PB \rightarrow \D \Prho^{0}$ and $\Bpm \rightarrow \D \Prho^{\pm}$ decays, labelled $\D \rho$, is described by the sum of a Gaussian and an exponential function.  
A Gaussian function is included for background candidates partially reconstructed from $\PB \rightarrow \Dstarmp \Prho^{\pm}$ and $\Bpm \rightarrow \Dstar \Prho^{\pm}$ decays.  
The background contribution from partially reconstructed \mbox{$\PB \rightarrow \D \Kstar$} decays is modelled by the convolution of an ARGUS function~\cite{Albrecht:1990am} with a Gaussian function;
the same convolution of functions is used for candidates reconstructed from \mbox{$\PB_s \rightarrow \D \Kstar$} decays.

For the fit to the \BpmtoDKspipipipm\ mass distribution, the same PDFs are used for signal, combinatorial 
and cross-feed background contributions as for the fit to the \BpmtoDKspipiKpm\ distribution.
The analogous function parameters are fixed to the results of fits to simulation. 
Again, functions are also included for partially reconstructed background candidates, with all parameters fixed to values obtained from simulation.  
The sum of two pairs of Gaussian functions, labelled $\Dstar \pi$, is used for the background from partially reconstructed $\Bpm \rightarrow \Dstar \pipm$ and
$\PB \rightarrow \Dstarmp \pipm$ decays.  
Partially reconstructed $\PB \rightarrow \D \Prho^{0}$ and $\Bpm \rightarrow \D \Prho^{\pm}$ decays are described by the convolution of an ARGUS function with a Gaussian function.  
A Gaussian function is used to describe background from partially reconstructed $\PB \rightarrow \Dstarmp \Prho^{\pm}$ and $\Bpm \rightarrow \Dstar \Prho^{\pm}$ decays.

In the simultaneous fit, the mean values of the signal functions in \BpmtoDKpm\ and \BpmtoDpipm\ are constrained to a common value.

The yield of the cross-feed component in the fit to the \BpmtoDKpm\ (\BpmtoDpipm) distribution is fixed
with respect to the signal yield in the \BpmtoDpipm\ (\BpmtoDKpm) distribution,
using knowledge of the efficiency and misidentification rate of the PID criterion separating the \BpmtoDKpm\ and \BpmtoDpipm\ candidate samples.
Large calibration samples of kaons and pions from $\Dstarpm \rightarrow \D(\rightarrow \Kmp \pipm) \pipm$ decays, kinematically selected from data,
 are reweighted to match the kinematic properties of the bachelor tracks in the \BpmtoDpipm\ {\it long} and {\it downstream} candidate samples
and are then used to determine the relevant efficiencies.
The remaining background yields are free to vary in the fit, as are the remaining PDF parameters and the ratio of the signal yields.

Since it is not possible to separate the two components of combinatorial
background with the fit to the \Bpm invariant mass distributions,
the yield of combinatorial \D background candidates is estimated from data 
using {\mbox{$\Bpm \rightarrow \D h^{\pm}$}} decays,
where the \D is reconstructed to decay to two same-sign pions ($\D \rightarrow \KS \pip \pip$ and charge conjugate).
These ``wrong-sign''  decays are subject to the selection criteria described in Sec.~\ref{sec:Selection}.

\subsection{\textbf{\textit{CP}} asymmetry fit}
\label{sec:DalitzFit}
The distributions in the \DtoKspipi\ decay phase space for positively and negatively charged \BpmtoDKspipiKpm\
and \BpmtoDKspipipipm\ candidate decays are fitted simultaneously using an unbinned maximum likelihood fit
to determine the \CP violation observables \xypm and the variation in efficiency over the phase space.
Although \BpmtoDKspipipipm\ decays are expected to exhibit interference analogous to \BpmtoDKspipiKpm\ decays and therefore be sensitive to $\gamma$,
the magnitude of the ratio of interfering \D\ decay amplitudes, $r_{\Bpm \rightarrow \D \pipm}$,
is expected to be an order of magnitude smaller than $r_{\B}$ for \BpmtoDKspipiKpm\ decays.
It is therefore possible, to a good approximation, to neglect the suppressed contribution to the \BpmtoDKspipipipm\ decay amplitude
and use \BpmtoDKspipipipm\ decays to obtain the efficiency variation as a function of $m^2_{+}$ and $m^2_{-}$, which is modelled as 
a second-order polynomial function.
This assumption is considered as a source of systematic uncertainty.

The candidates are divided into eight subsamples, according to \KS\ type ({\it long} or {\it downstream}), the
charge of the bachelor track, and whether the candidate is identified as a \BpmtoDKpm\ or \BpmtoDpipm\ decay.
The negative logarithm of the likelihood,
\begin{equation}
  - \ln \mathcal{L} = - \sum_{s} \sum_{k} \ln \left(
                \sum_{c} N_{c} \cdot p_{cs}^\mathrm{mass} \left( \left( m_{Dh} \right)_{k} ; \vec{P}_{cs}^\mathrm{mass}  \right)
                \cdot p_{cs}^\mathrm{model}\left( \left( m^2_{+}, m^2_{-} \right)_{k}; \vec{P}_{cs}^\mathrm{model} \right) \right),
\end{equation}
is minimised; 
in this expression, $c$ indexes the candidate categories (signal or background type), $s$ indexes the subsample,
and $k$ identifies each decay candidate.
$N_{c}$ is the candidate yield for category $c$,
and
$p_{cs}^\mathrm{mass}$ is the invariant mass PDF,
$p_{cs}^\mathrm{model}$ is the normalised \D\ decay model described below,
$\vec{P}_{cs}^\mathrm{mass}$ are the mass PDF parameters, and
$\vec{P}_{cs}^\mathrm{model}$ are the \D\ decay model parameters for category $c$ and subsample $s$.
It should be noted that \xypm are included in the parameter list of the \BpmtoDKpm\ signal category and the \BpmtoDpipm\ cross-feed category, 
which arises from misidentification of \BpmtoDKpm\ decays.
The normalisation of $p_{cs}^\mathrm{model}$ depends on the efficiency variation over the
phase space.
The yields and parameters of the mass PDFs are fixed to the results obtained in the \Bpm invariant mass fit.
To avoid inadvertent experimenter's bias in the determination of the \CP violation parameters, 
the values of the observables \xypm are masked until the measurement technique has been finalised.

The model describing the amplitude of the \DtoKspipi\ decay over the phase space, $A_f\left(m_+^2, m_-^2 \right)$, is identical to
that used by the \babar collaboration in Refs.~\cite{delAmoSanchez:2010rq, delAmoSanchez:2010xz}.
It incorporates an isobar model
for P-wave (which includes $\Prho(770)$, $\Pomega(782)$, Cabibbo-allowed and doubly Cabibbo-suppressed $\Kstar (892)$ and $\Kstar (1680)$) 
and D-wave (including $f_2(1270)$ and $\Kstar_2 (1430)$)
contributions. 
A generalised LASS amplitude for the $\kaon \pi$ S-wave contribution ($\Kstar_0 (1430)$) and a K-matrix
with P-vector approach for the $\pi \pi$ S-wave contribution are also included in the model.
All parameters of the model are fixed in the fit to the values determined in Ref.~\cite{delAmoSanchez:2010xz}.
\footnote{The model implemented by BaBar~\cite{delAmoSanchez:2010xz} differs from the formulation described
  therein. One of the two Blatt-Weisskopf coefficients was set to unity, and the imaginary part of the
  denominator of the Gounaris-Sakurai propagator used the mass of the resonant pair, instead of the mass
  associated with the resonance. The model used herein replicates these features without modification. It has
  been verified that changing the model to use an additional centrifugal barrier term and a modified
  Gounaris-Sakurai propagator has a negligible effect on the measurements.}

The fit is performed using refitted candidates with a \Bpm invariant mass lying within $\pm 50\mevcc$
around the known value~\cite{PDG2012},
corresponding to an invariant mass region of approximately $\pm 3 \sigma$ around the signal peak.
Although the full description of the mass PDF provides valuable constraints for the background 
within the mass window, only those backgrounds with significant contributions are included in the \CP asymmetry fit.
The yields of the signal and incorporated background contributions are given in Table~\ref{tab:SigWindowYields}.
For the \BpmtoDKpm\ subsamples, the cross-feed, combinatorial \D, random $Dh$, $\Dstar \pi$, $\D \rho$ and
\mbox{$\PB_s \rightarrow \D \Kstar$} background categories are included in the fit.
The cross-feed contribution is assumed to be distributed in the phase space of the \DtoKspipi\ decay according
to the \DztoKspipi\ (\DzbtoKspipi ) decay model in the \BmtoDKm\ (\BptoDKp ) case.
Combinatorial \D background candidates are expected to be distributed non-resonantly over the phase space.
The distribution of random $Dh$ candidates is assumed to be an incoherent sum of the \DztoKspipi\ and \DzbtoKspipi\ decay models.
Both $\Bpm \rightarrow \Dstar \pipm$ and $\PB \rightarrow \Dstarmp \pipm$ decays are represented
by the inclusion of a \DztoKspipi\ (\DzbtoKspipi ) decay model in the \BmtoDKm\ (\BptoDKp ) case.
The $\D \rho$ component of the invariant mass fit is composed of candidates from $\PB \rightarrow \D \Prho^{0}$
and $\Bpm \rightarrow \D \Prho^{\pm}$ decays; the distribution of candidates from $\PB \rightarrow \D \Prho^{0}$ over the
\DtoKspipi\ decay phase space is assumed to be an incoherent sum of the \DztoKspipi\ and \DzbtoKspipi\ decay models, whereas the
candidates from $\Bpm \rightarrow \D \Prho^{\pm}$ are accounted for with a \DztoKspipi\ (\DzbtoKspipi ) decay model for the \BmtoDKm\ (\BptoDKp ) case.
Background \mbox{$\PB_s \rightarrow \D \Kstar$} candidates are assumed to be distributed according to the \DzbtoKspipi\
(\DztoKspipi ) decay model in the \BmtoDKm\ (\BptoDKp ) case.
For the \BpmtoDpipm\ subsamples, contributions from
cross-feed, combinatorial \D, random $Dh$, and $\Dstar \pi$ background types are included in the fit.
The cross-feed candidates in \BpmtoDpipm\ arise from misidentification of the bachelor track 
of \BpmtoDKpm\ decays; the candidates are assumed to be distributed accordingly.
The remaining combinatorial and $\Dstar \pi$ background contributions are assumed to be distributed as described above.

\begin{table}[t]
\begin{small}
  \setlength{\tabcolsep}{0pt}
  \caption{
    \small Signal and background yields for components contributing to the \CP asymmetry fit, in the region $\pm 50\mevcc$ around the known \Bpm meson mass.}
    \begin{center}\begin{tabular}{llrlrl}
    \hline
    \mcol{2}{Fit component}                   & \mcol{2}{\BpmtoDKpm, {\it long}~~~~~}  & \mcol{2}{\BpmtoDKpm, {\it downstream}}  \\ 
    \hline
    \mcol{2}{Signal}                          & ~~~~~~~$ 217  $&$\pm 17 $    & ~~~~~~~~~~$ 420 $&$ \pm 27 $         \\
    \mcol{2}{Backgrounds}		      &                &             &                  &                   \\
     ~~~~& Cross-feed (from \BpmtoDpipm)~~~   & $ 35.9 $&$\pm 0.7 $       & $ 76  $&$ \pm 1 $          \\
     & Combinatorial \D                       & $ 5    $&\hspace{0.042cm}$^{+\hspace{0.02cm}7}_{-\hspace{0.02cm}3} $    & $ 31  $&\hspace{0.042cm}$^{+\hspace{0.02cm}11}_{-\hspace{0.02cm}9} $     \\
     & Random $Dh$                            & $ 28   $&\hspace{0.042cm}$^{+\hspace{0.02cm}5}_{-\hspace{0.02cm}8} $    & $ 45  $&\hspace{0.042cm}$^{+\hspace{0.02cm}18}_{-\hspace{0.02cm}19} $    \\
     & $\Dstar \pi$                           & $ 0.36 $&$ \pm 0.08 $     & $ 6   $&$ \pm 7 $          \\
     & $\D \rho$                              & $ 2.2  $&$ \pm 0.5 $      & $ 4   $&$ \pm 11 $         \\
     & $\PB_s \rightarrow \D \Kstar$          & $ 0.9  $&$ \pm 0.2 $      & $ 4   $&$ \pm 2 $          \\
    \hline
    \\
    \hline
    \mcol{2}{Fit component}                   & \mcol{2}{\BpmtoDpipm, {\it long}}   & \mcol{2}{\BpmtoDpipm, {\it downstream}}  \\
    \hline
    \mcol{2}{Signal}                          & $ 2906 $&$ \pm 56 $       & $ 5960 $&$ \pm 80 $        \\
    \mcol{2}{Backgrounds}		      &         &                 &         &                  \\
     & Cross-feed (from \BpmtoDKpm)~~~        & $ 27   $&$\pm 2 $         & $ 53   $&$\pm 3 $          \\
     & Combinatorial \D                       & $ 15   $&\hspace{0.042cm}$^{+\hspace{0.02cm}19}_{-\hspace{0.02cm}10} $  & $ 99   $&\hspace{0.042cm}$^{+\hspace{0.02cm}36}_{-\hspace{0.02cm}27} $   \\
     & Random $Dh$                            & $ 76   $&\hspace{0.042cm}$^{+\hspace{0.02cm}15}_{-\hspace{0.02cm}22} $  & $ 146  $&\hspace{0.042cm}$^{+\hspace{0.02cm}33}_{-\hspace{0.02cm}41} $   \\
     & $\Dstar \pi$                           & $ 6.6  $&$\pm 0.4 $       & $ 22.0 $&$\pm 0.7 $        \\
    \hline
  \end{tabular}\end{center}
\label{tab:SigWindowYields}
\end{small}
\end{table}

\begin{figure}[h]
  \begin{center}
   \hspace{0.04\textwidth}
    \includegraphics[width=0.35\textwidth]{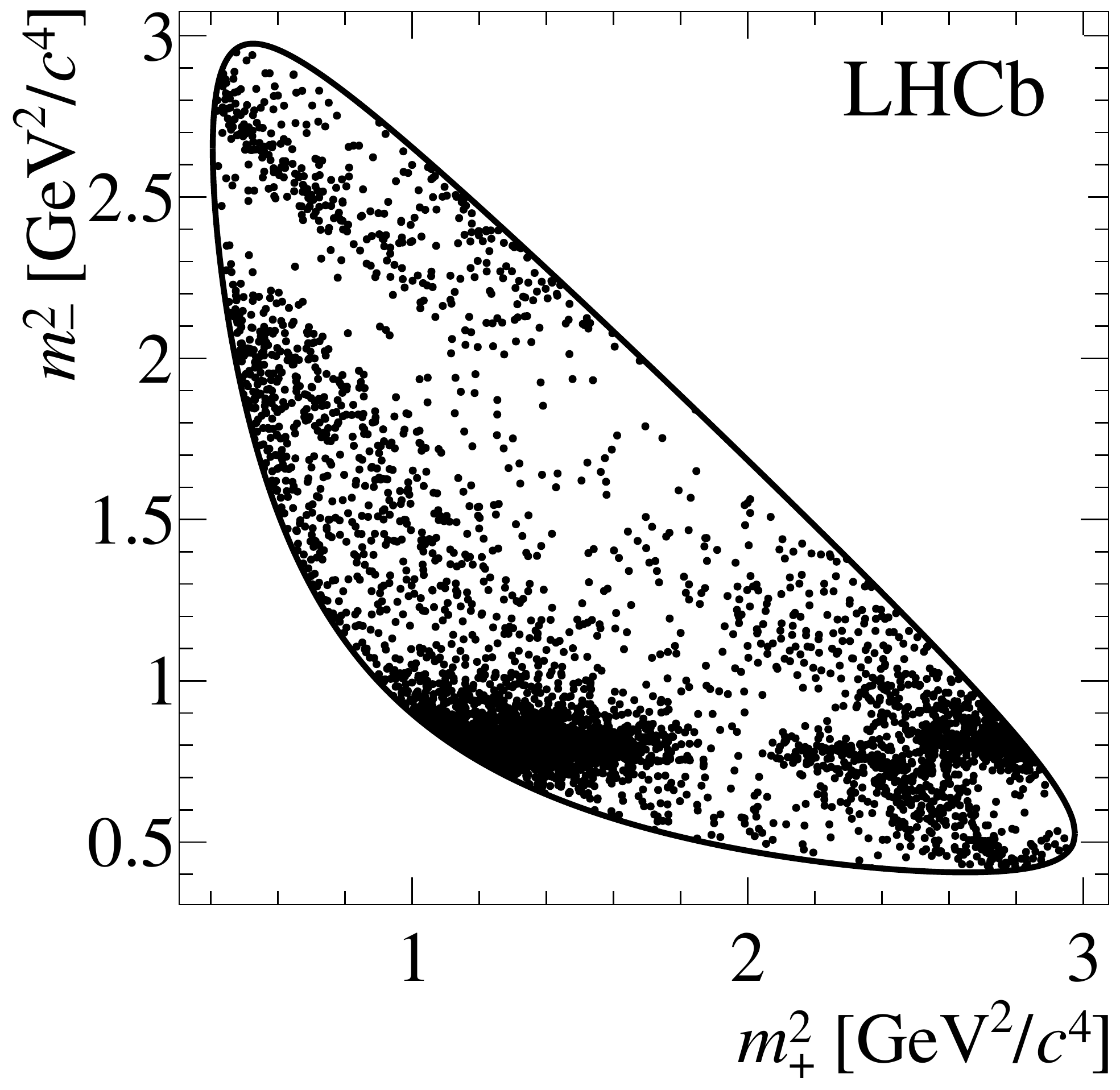} \hspace{0.07\textwidth}
    \includegraphics[width=0.48\textwidth]{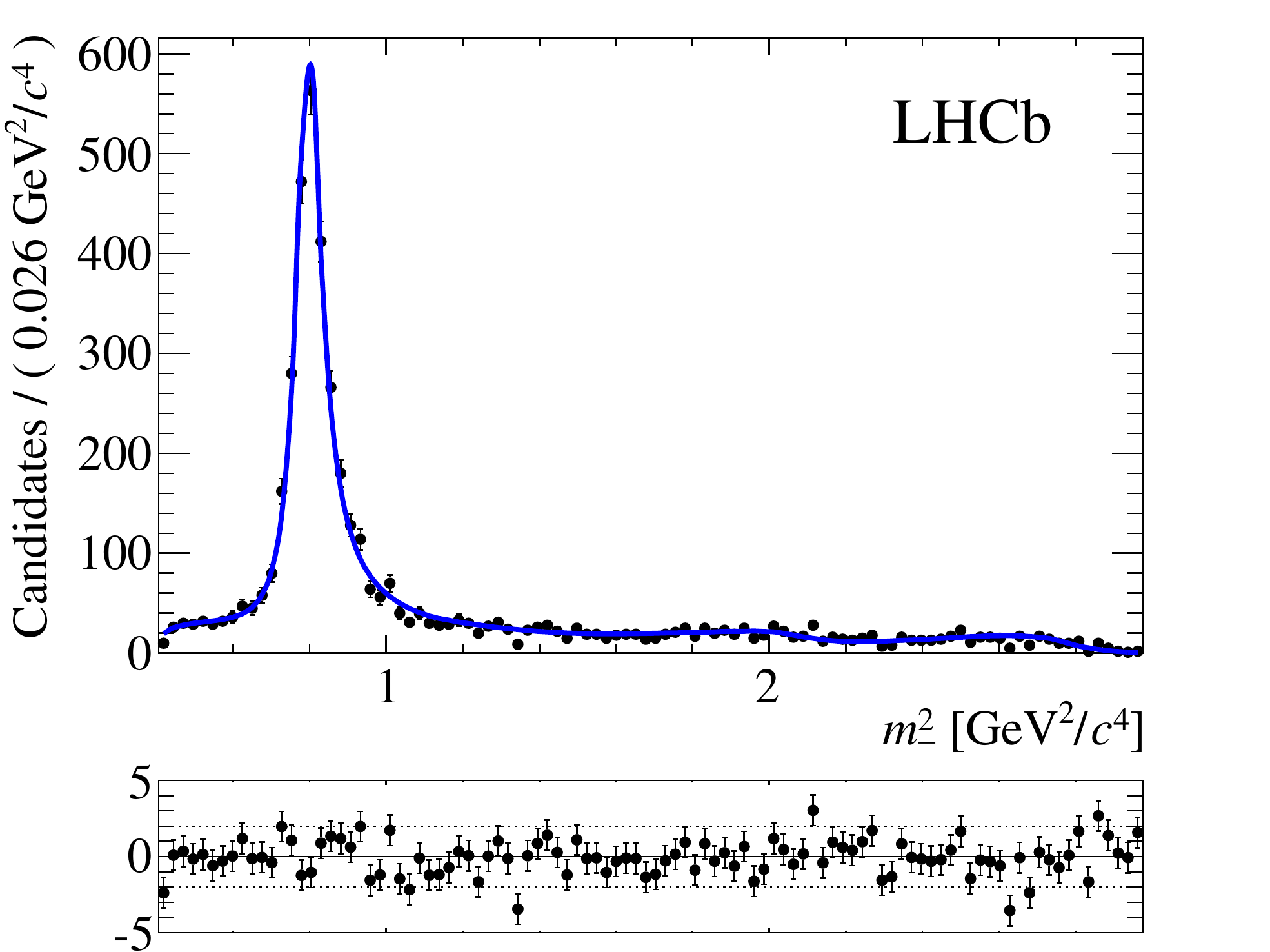}
    \includegraphics[width=0.48\textwidth]{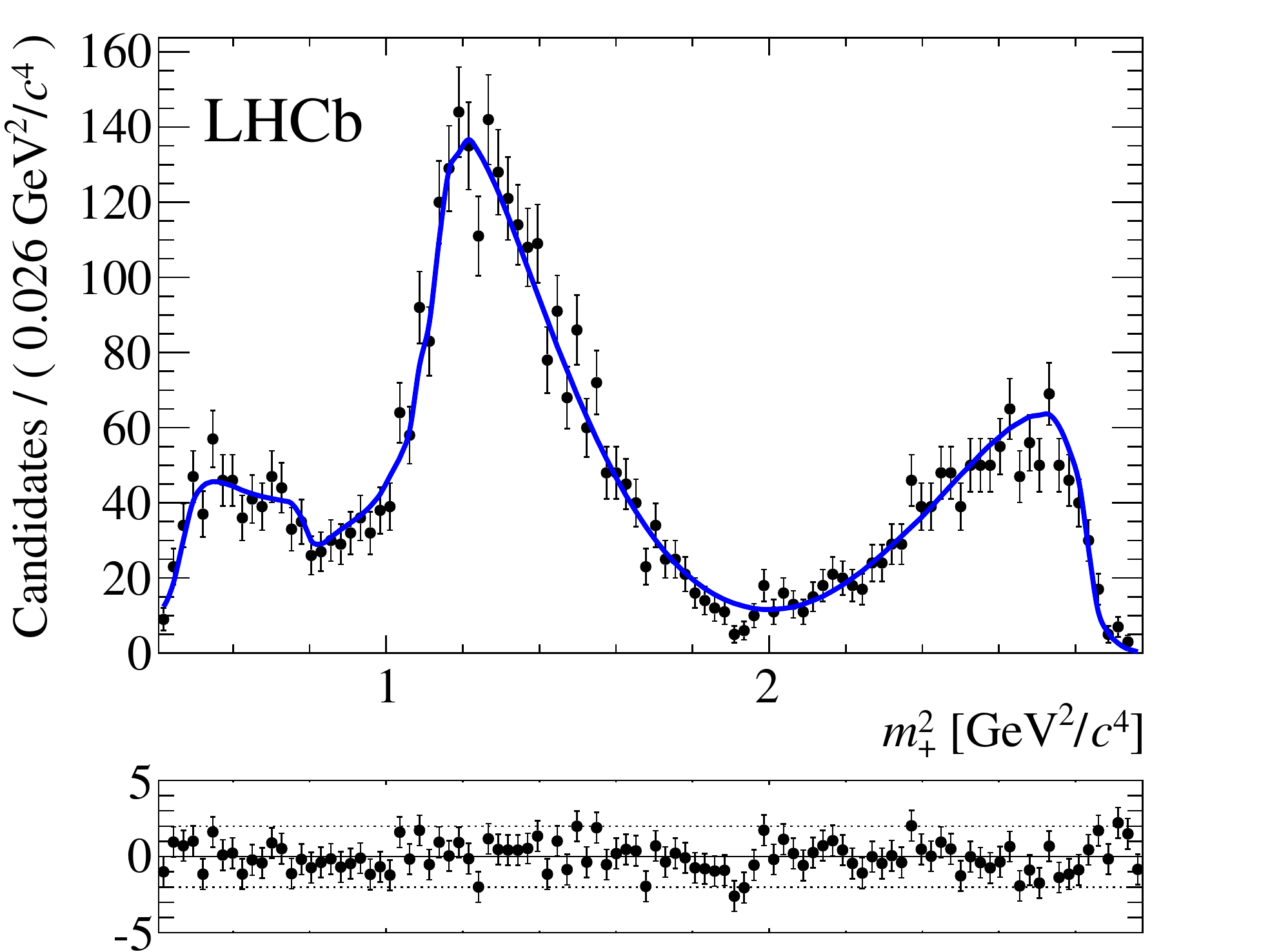}
    \includegraphics[width=0.48\textwidth]{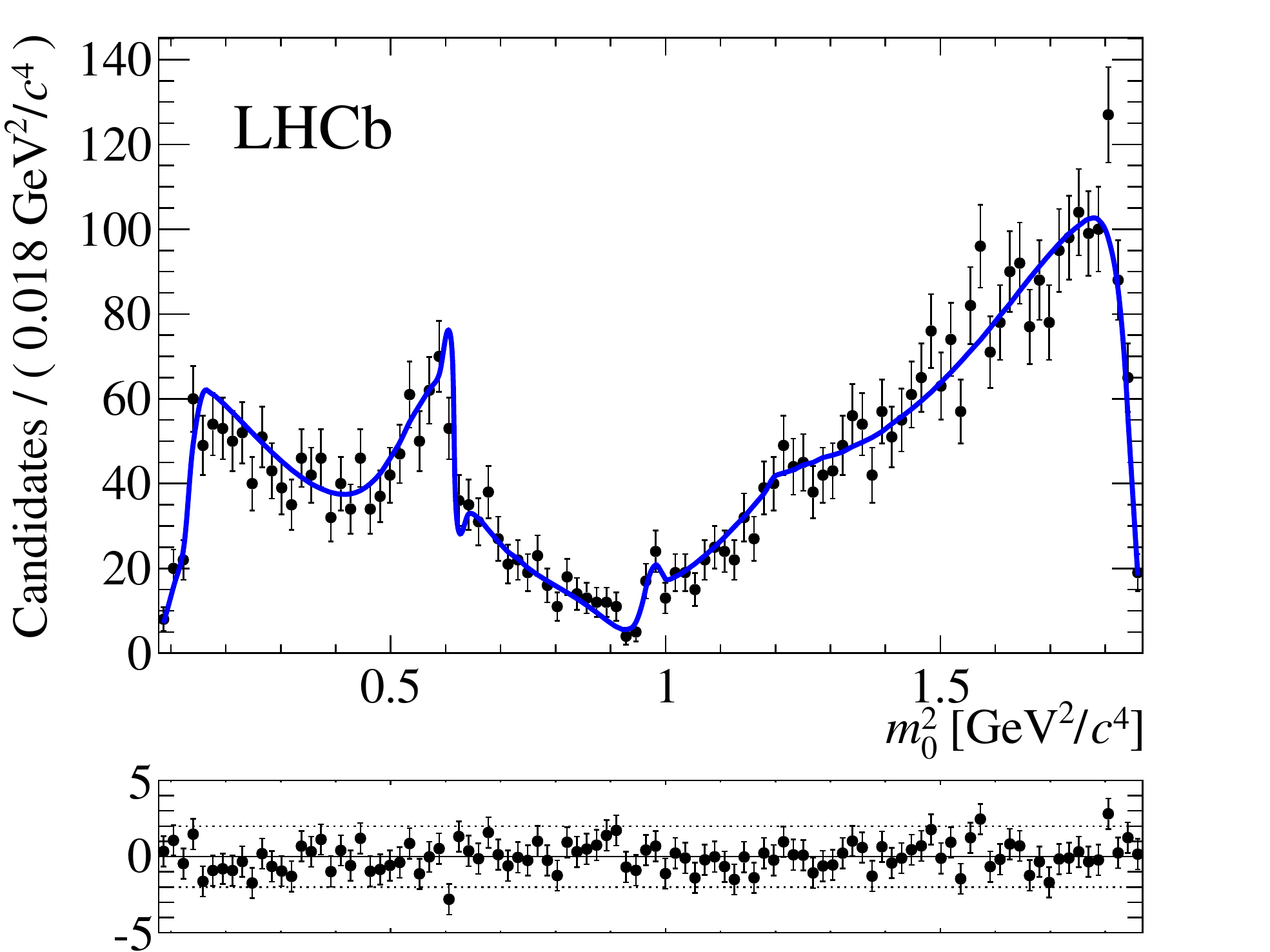}
    \caption{Dalitz plot and its projections, with fit result superimposed, for $B^- \rightarrow D \pi^-$ candidates; \mbox{$m_{\pm}^2 \equiv m_{\KS \pipm}^2$}
    and \mbox{$m_{0}^2 \equiv m_{\pip \pim}^2$}. The lower parts of the figures are normalised residual distributions.}
    \label{fg.fit.projDpim}
  \end{center}
\end{figure}

\begin{figure}[h]
  \begin{center}
   \hspace{0.04\textwidth}
    \includegraphics[width=0.35\textwidth]{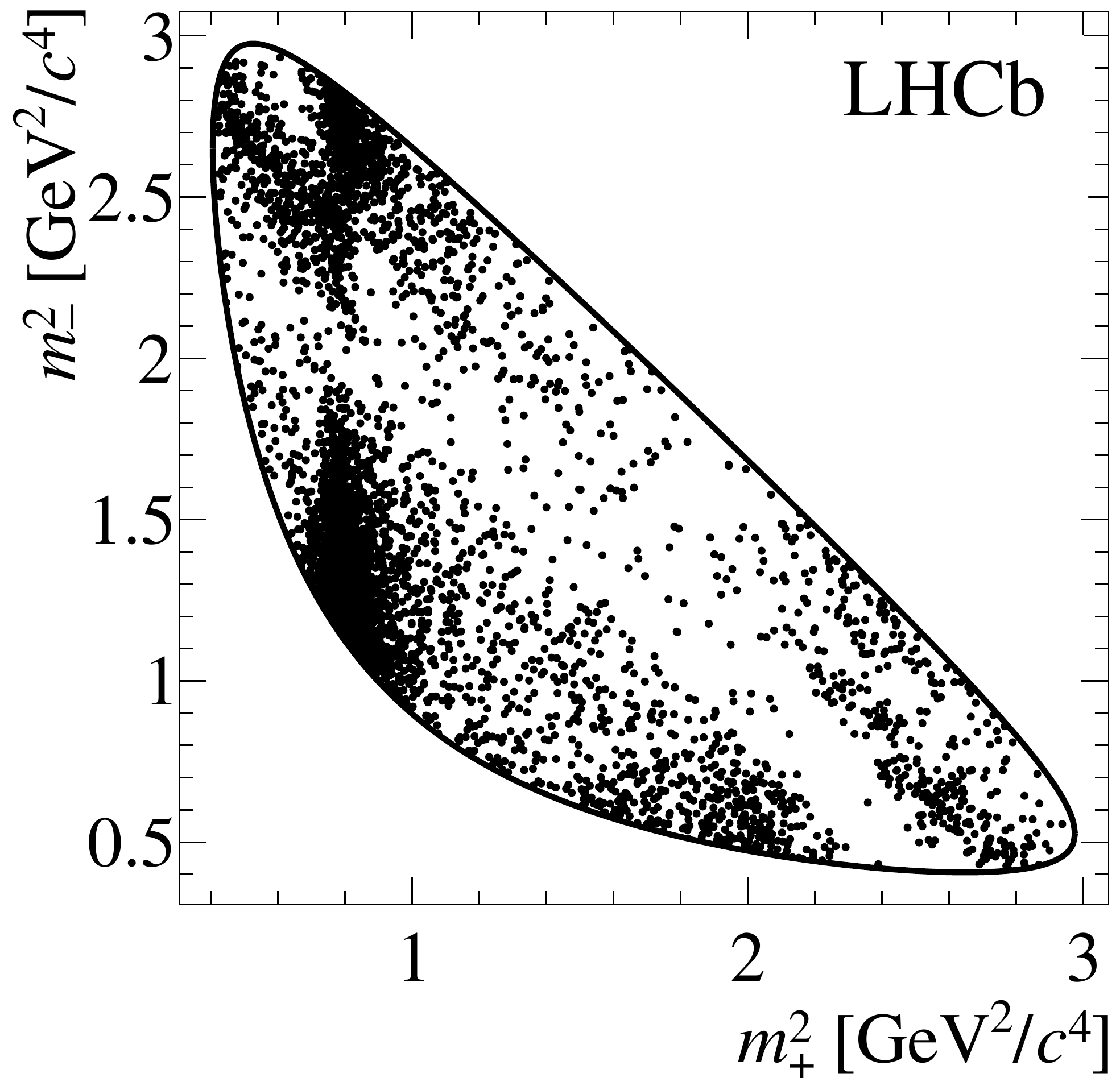} \hspace{0.07\textwidth}
    \includegraphics[width=0.48\textwidth]{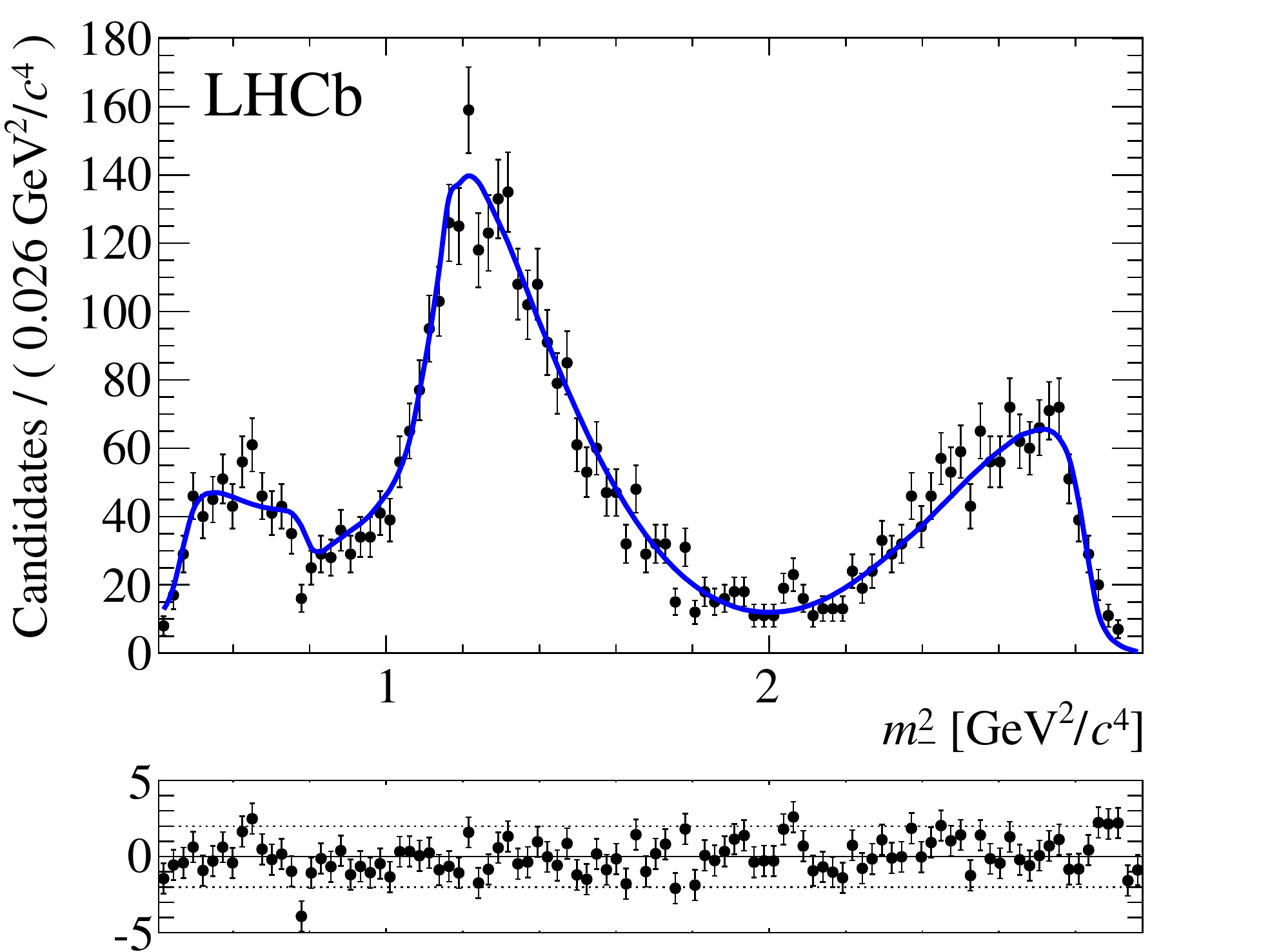}
    \includegraphics[width=0.48\textwidth]{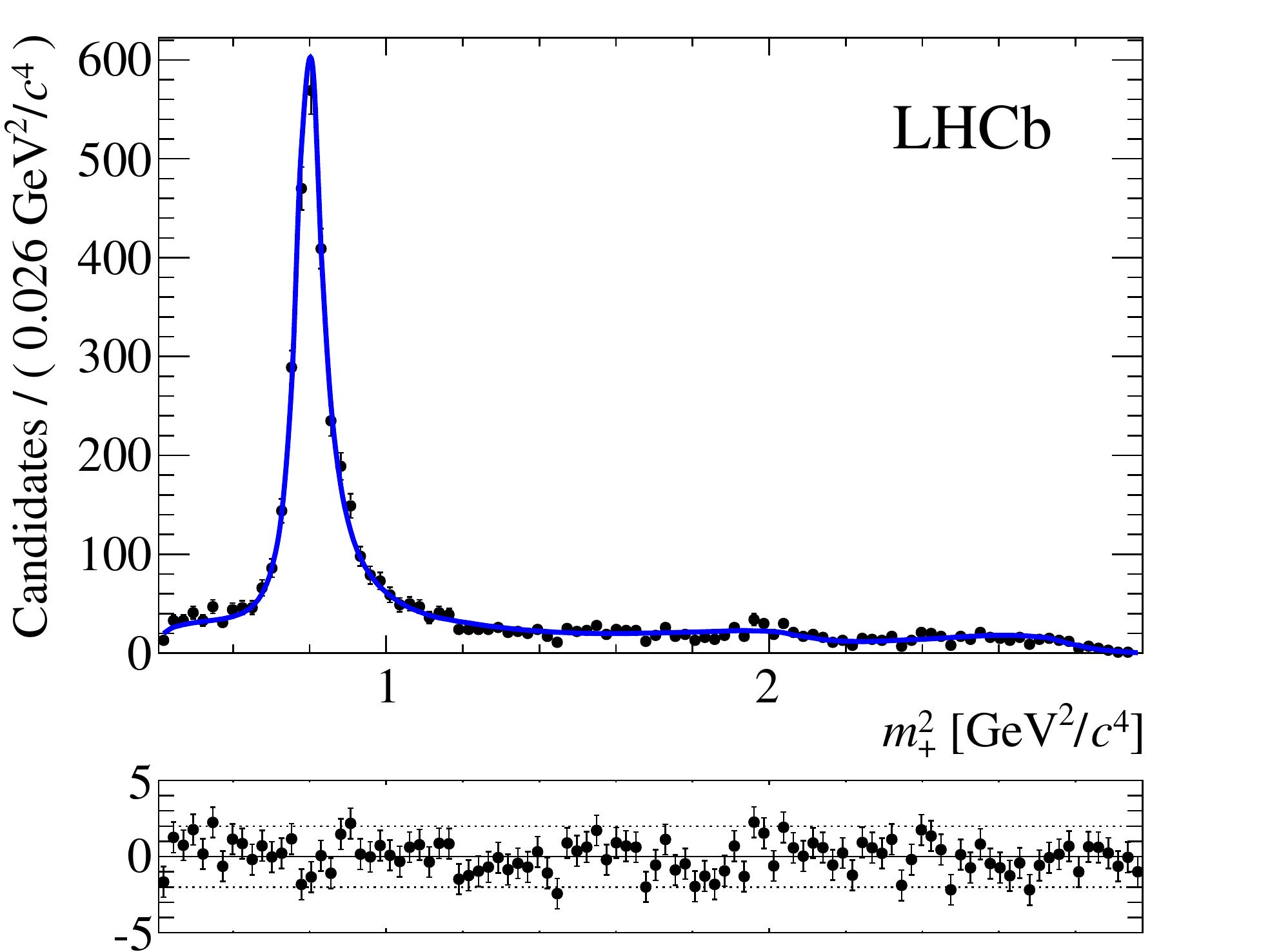}
    \includegraphics[width=0.48\textwidth]{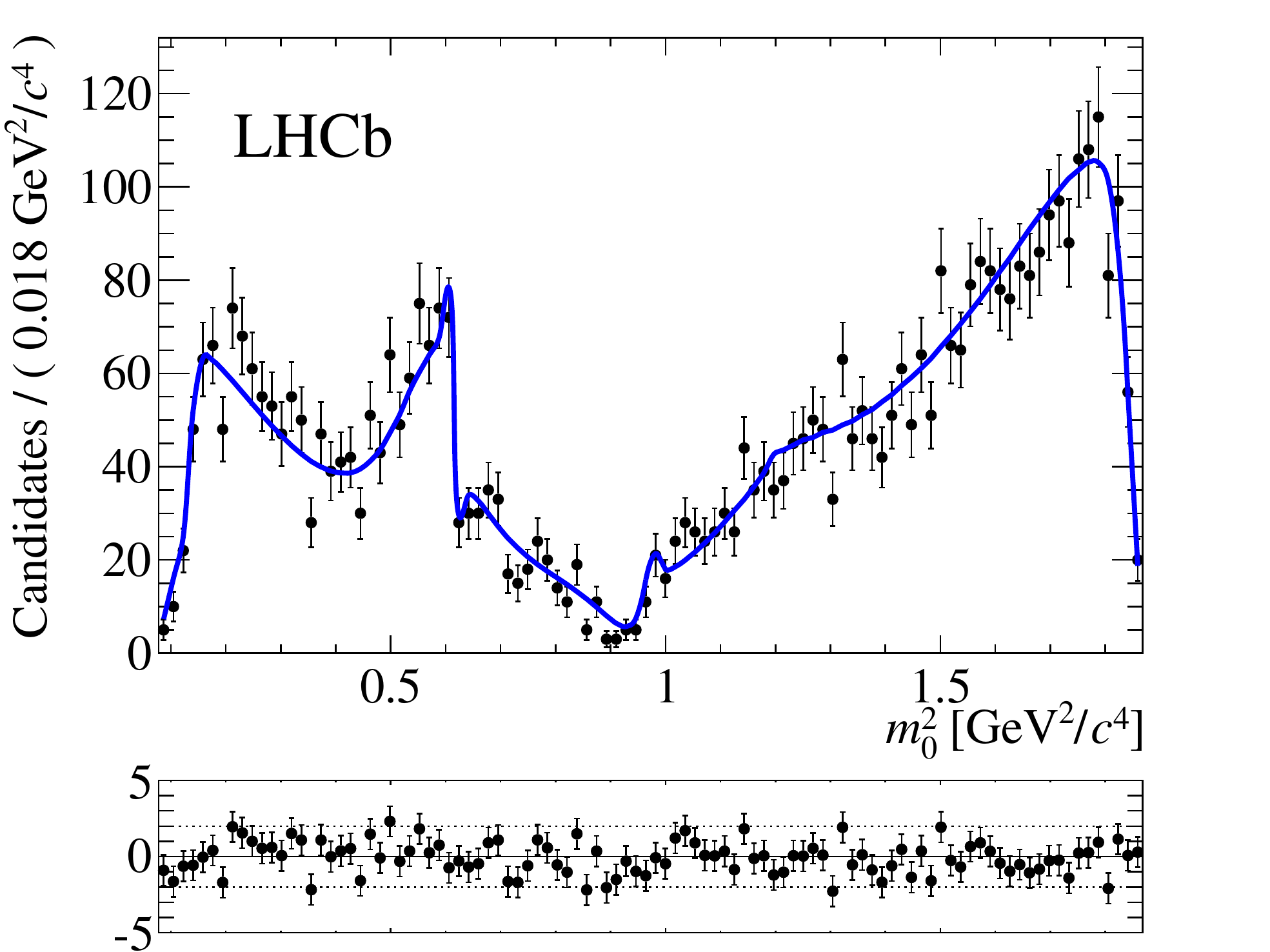}
    \caption{Dalitz plot and its projections, with fit result superimposed, for $B^+ \rightarrow D \pi^+$ candidates; \mbox{$m_{\pm}^2 \equiv m_{\KS \pipm}^2$}
    and \mbox{$m_{0}^2 \equiv m_{\pip \pim}^2$}. The lower parts of the figures are normalised residual distributions.}
    \label{fg.fit.projDpip}
  \end{center}
\end{figure}

\begin{figure}[h]
  \begin{center}
   \hspace{0.04\textwidth}
    \includegraphics[width=0.35\textwidth]{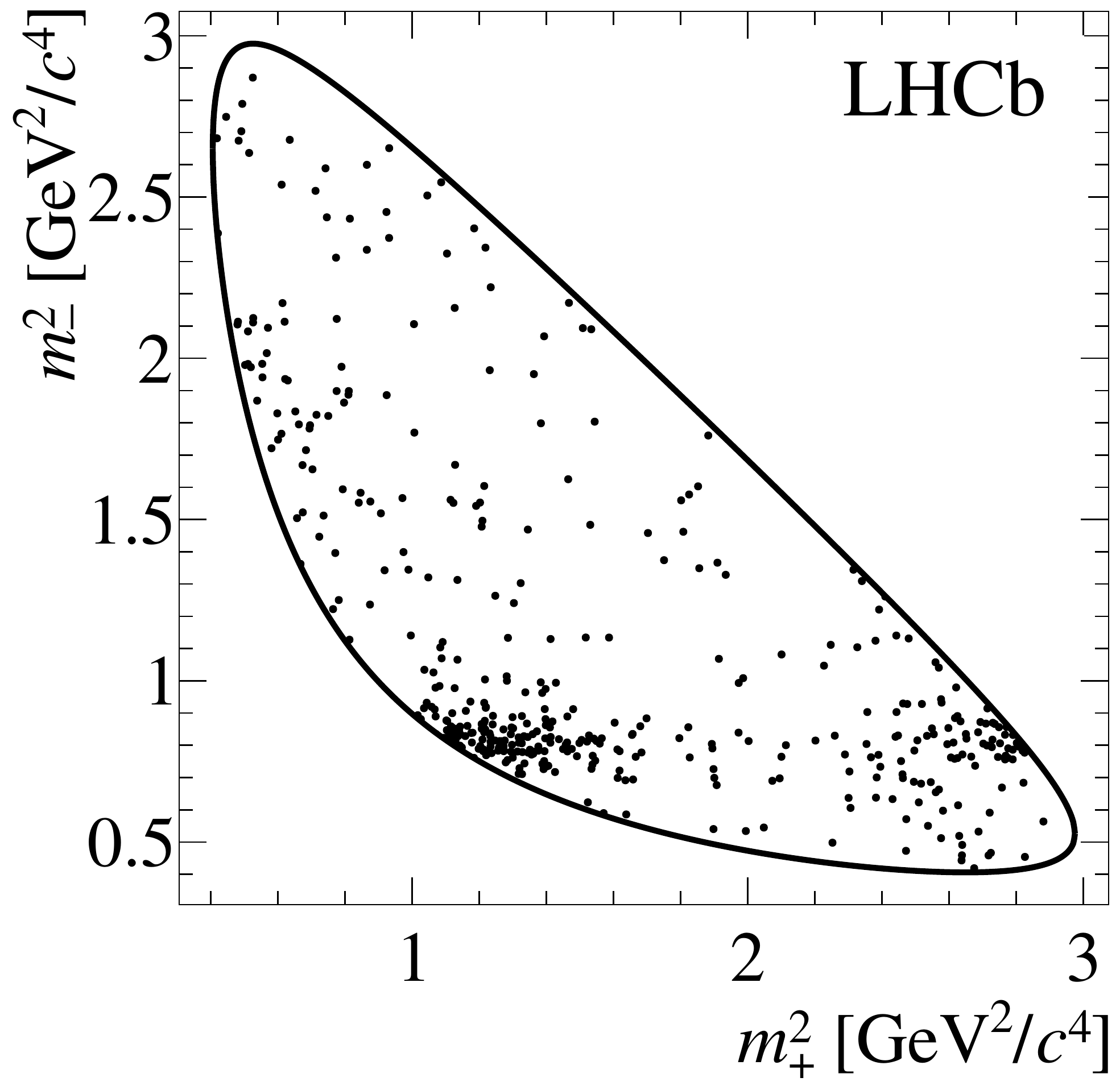} \hspace{0.07\textwidth}
    \includegraphics[width=0.48\textwidth]{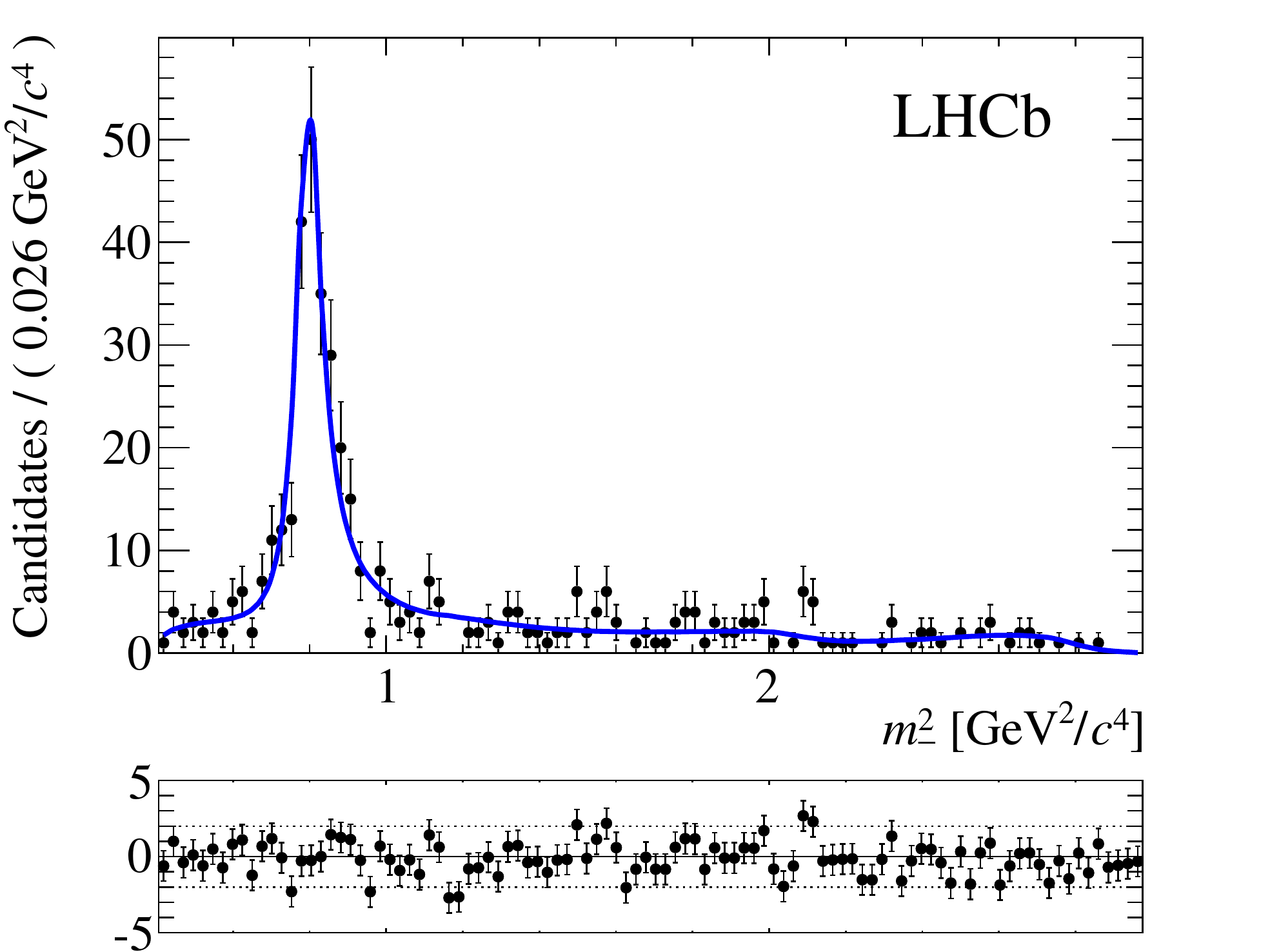}
    \includegraphics[width=0.48\textwidth]{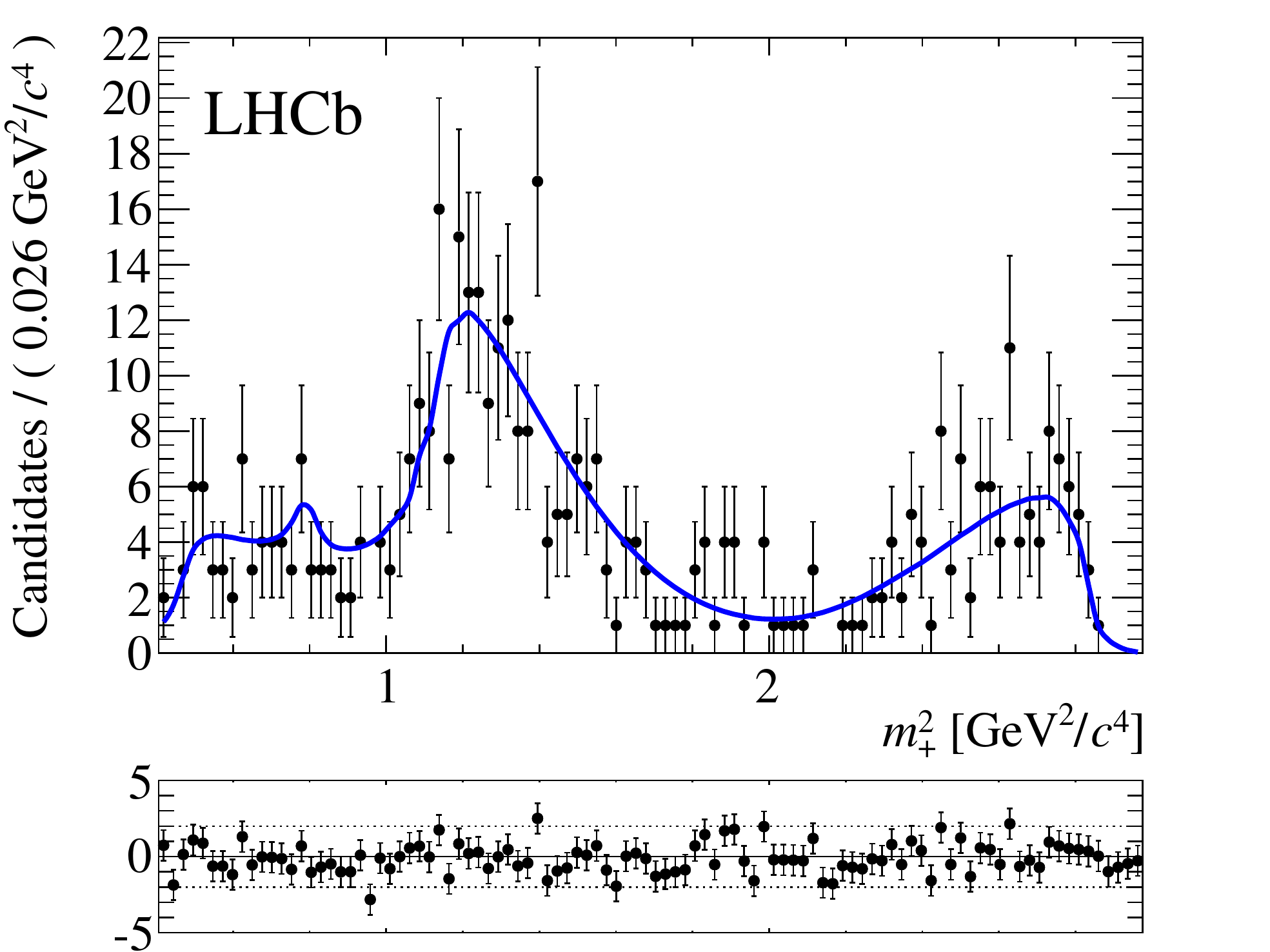}
    \includegraphics[width=0.48\textwidth]{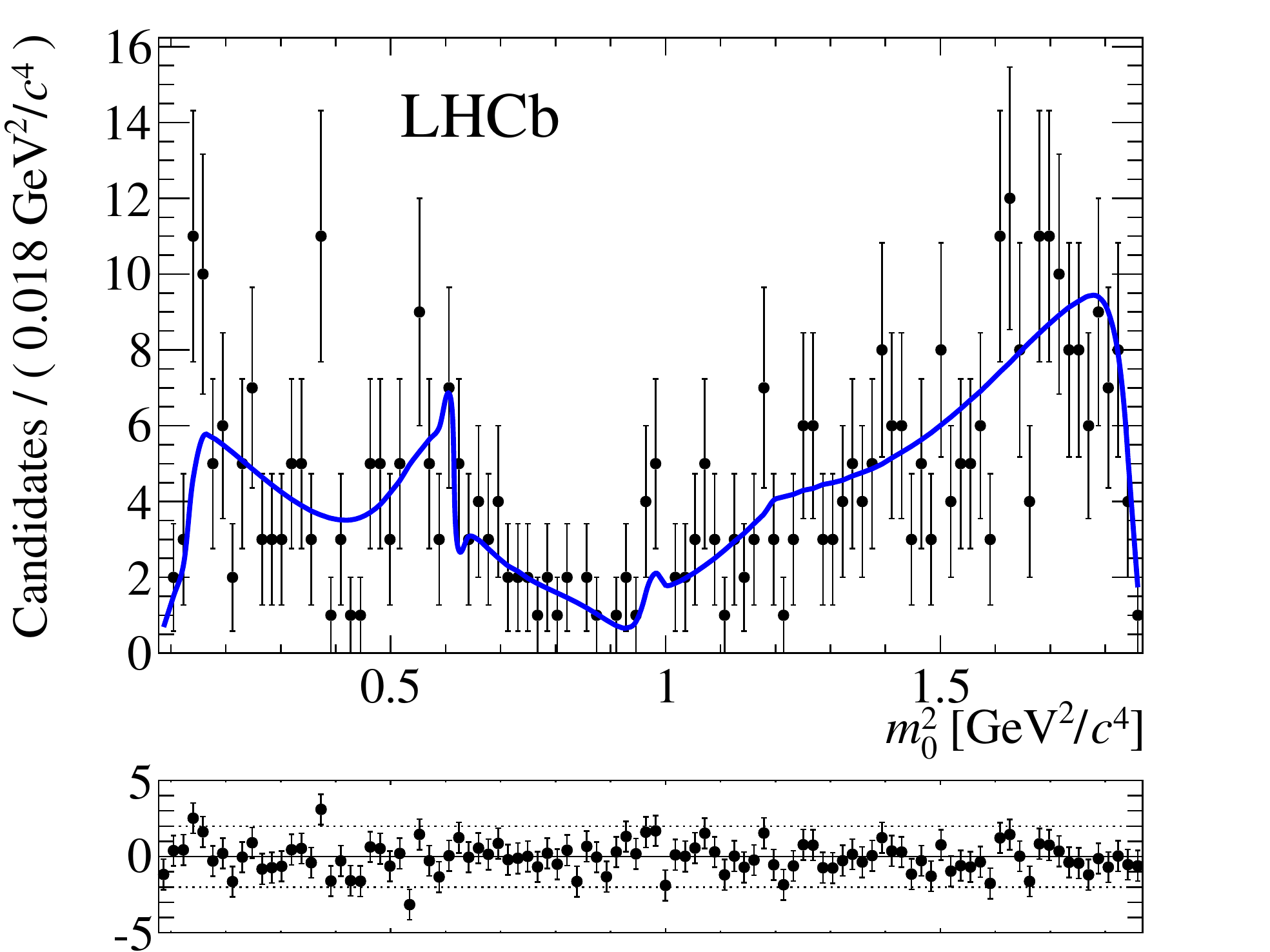}
    \caption{Dalitz plot and its projections, with fit result superimposed, for $B^- \rightarrow D K^-$ candidates; \mbox{$m_{\pm}^2 \equiv m_{\KS \pipm}^2$}
    and \mbox{$m_{0}^2 \equiv m_{\pip \pim}^2$}. The lower parts of the figures are normalised residual distributions.}
    \label{fg.fit.projDKm}
  \end{center}
\end{figure}

\begin{figure}[h]
  \begin{center}
   \hspace{0.04\textwidth}
    \includegraphics[width=0.35\textwidth]{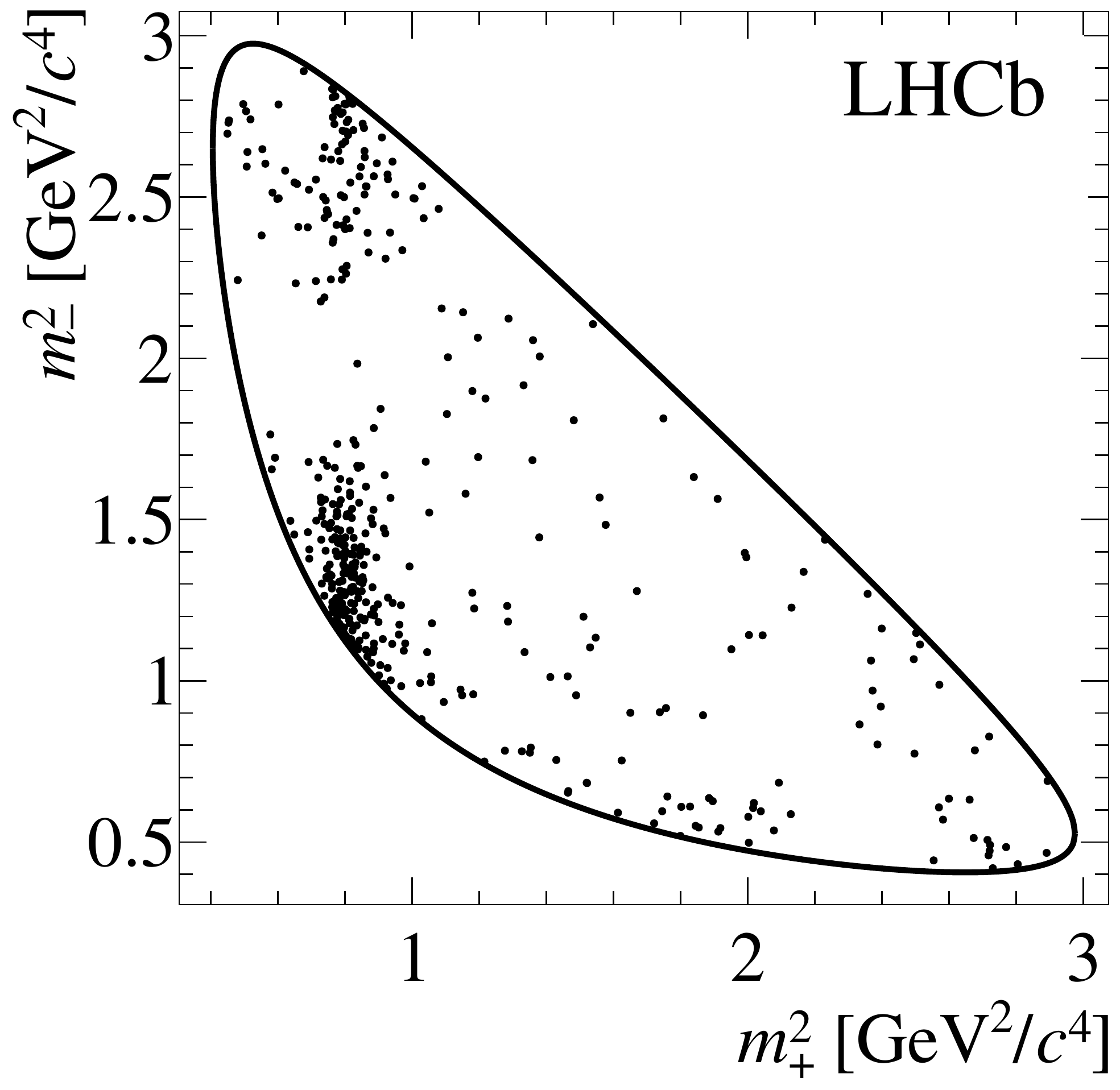} \hspace{0.07\textwidth}
    \includegraphics[width=0.48\textwidth]{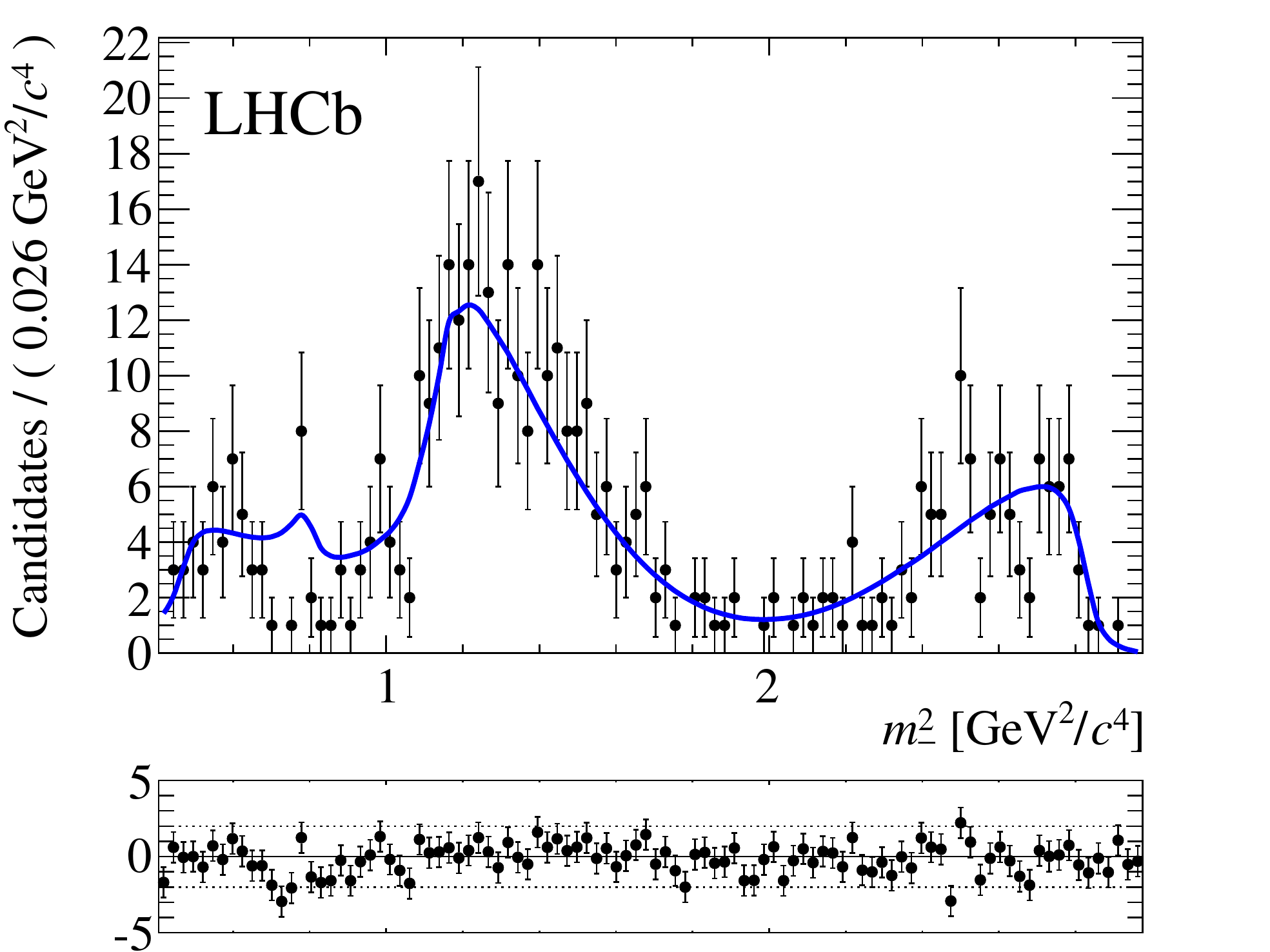}
    \includegraphics[width=0.48\textwidth]{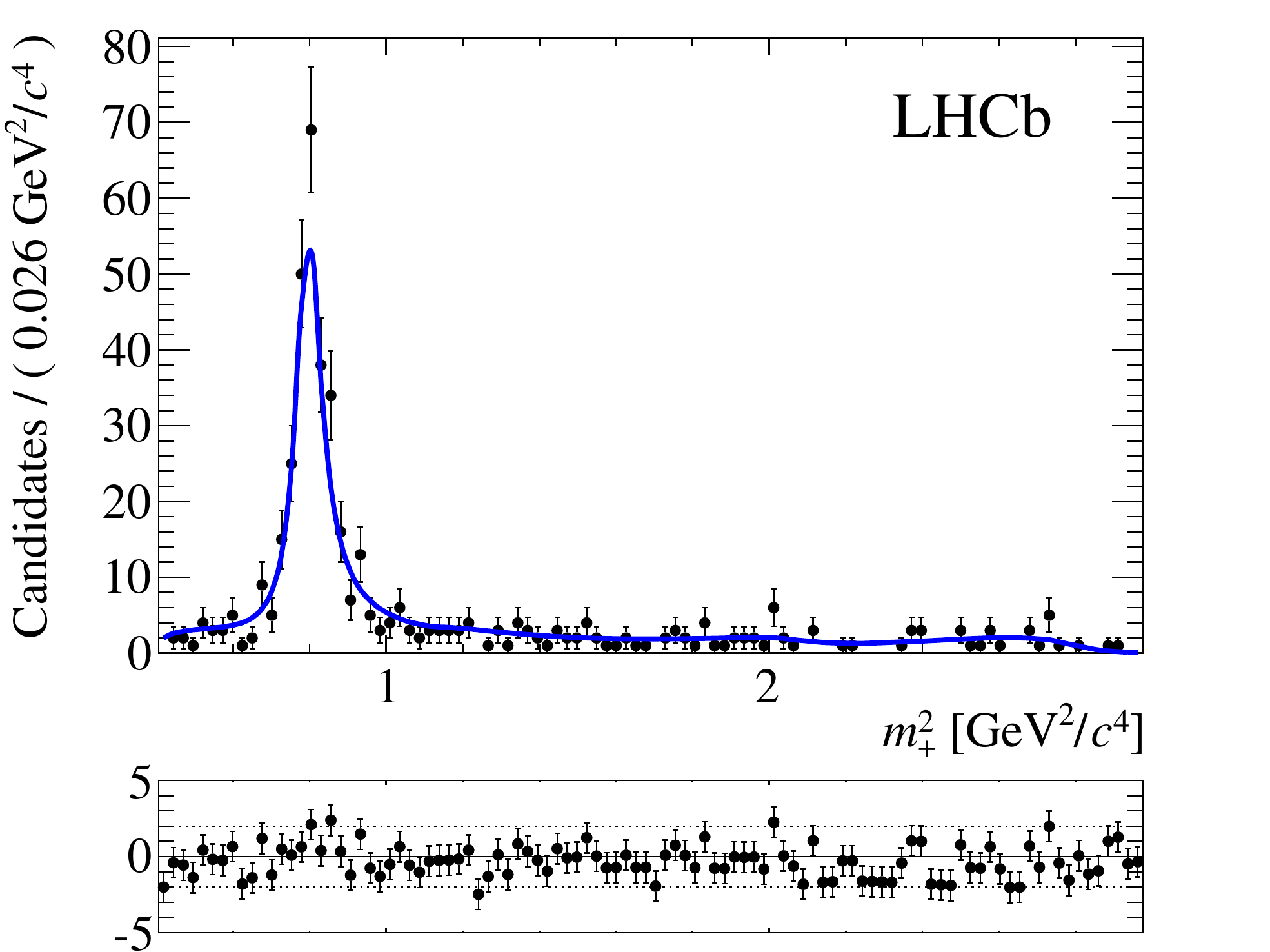}
    \includegraphics[width=0.48\textwidth]{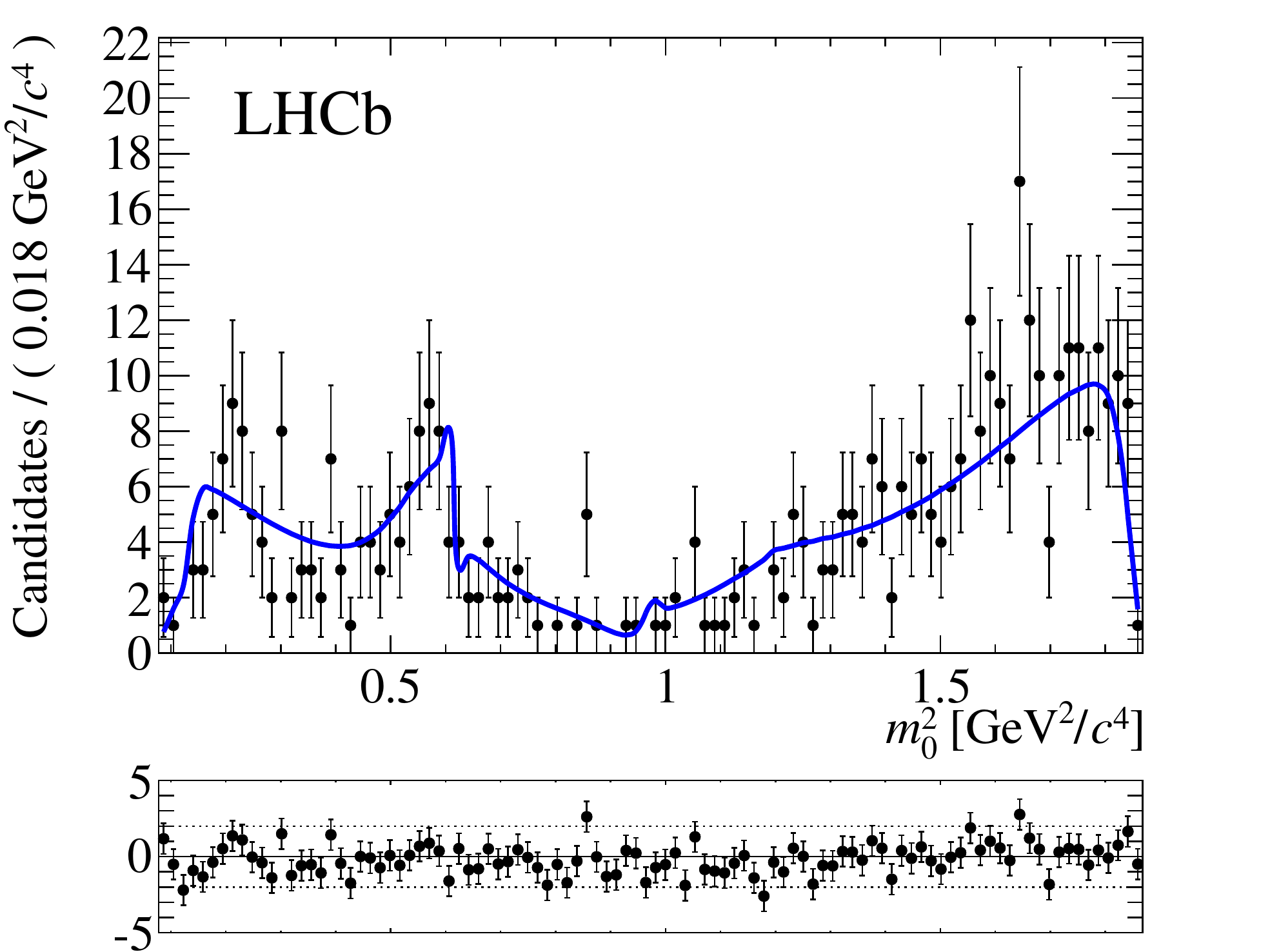}
    \caption{Dalitz plot and its projections, with fit result superimposed, for $B^+ \rightarrow D K^+$ candidates; \mbox{$m_{\pm}^2 \equiv m_{\KS \pipm}^2$}
    and \mbox{$m_{0}^2 \equiv m_{\pip \pim}^2$}. The lower parts of the figures are normalised residual distributions.}
    \label{fg.fit.projDKp}
  \end{center}
\end{figure}

\begin{figure}[h]
  \begin{center}
    \includegraphics[width=0.70\textwidth]{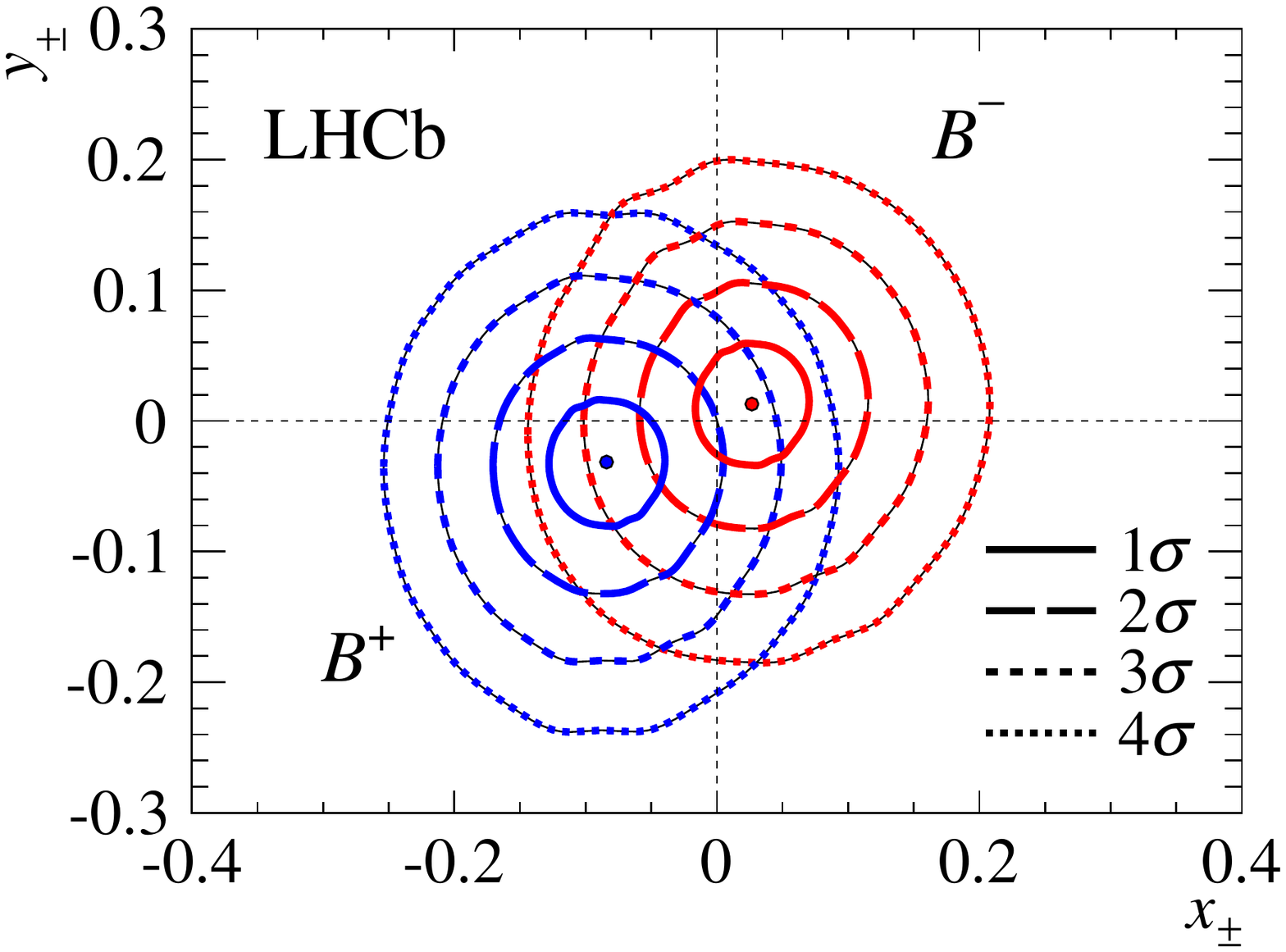}
    \caption{Likelihood contours at 39.35\%, 86.47\%, 98.89\% and 99.97\% confidence level for
      $(x_{+},y_{+})$\xspace (blue) and $(x_{-},y_{-})$\xspace(red).}
    \label{fg.fit.contours}
  \end{center}
\end{figure}

Figures \ref{fg.fit.projDpim}--\ref{fg.fit.projDKp} show the \BpmtoDKspipipipm\ and \BpmtoDKspipiKpm\ candidate Dalitz plot distributions and their projections,
with the results of the fit superimposed.
The resulting measured values of \xypm are
\begin{align*}
  x_- &= +0.027 \pm 0.044,  \\
  y_- &= +0.013 \pm 0.048,  \\
  x_+ &= -0.084 \pm 0.045,  \\
  y_+ &= -0.032 \pm 0.048,  
\end{align*}
where the uncertainties are statistical only.  The corresponding likelihood contours are shown in Fig. \ref{fg.fit.contours}.

\section{Systematic uncertainties}
\label{sec:Systematics}
Systematic uncertainties on the measured values of \xypm arising from various sources are considered and summarised in Table~\ref{tab:syst:summary}.
Unless otherwise stated, for each source considered the \CP asymmetry fit is repeated 
with the efficiency parameters and \xypm allowed to vary, as in the nominal fit to data.
The resulting differences in the values of \xypm from the nominal results are taken as systematic uncertainties.

\newcommand{\tmthree}{(\times 10^{-3})}

\begin{table}[tb]
\begin{small}
\caption{\small Absolute values of systematic uncertainties.  The \CP asymmetry fit bias is considered as a one-sided uncertainty and is included in the quadrature sum on that side only.}
\begin{center}
\begin{tabular}{llrrrr}
\hline
\mcol{2}{Source}                                          & $\delta x_{-}\tmthree$   & $\delta y_{-}\tmthree$   & $\delta x_{+}\tmthree$   & $\delta y_{+}\tmthree$   \\
\hline
\mcol{2}{Background yields}                               &                          &                          &                          &                          \\
& Cross-feed                                              & $0.21$                   & $0.96$                   & $0.65$                   & $0.26$                   \\
& Total combinatorial                                     & $1.1 $                   & $3.5 $                   & $1.7$                    & $2.7$                    \\
& Combinatorial \D                                        & $1.0 $                   & $4.3 $                   & $2.7$                    & $4.9$                    \\
\mcol{2}{Inclusion of semileptonic background}            & $3.1 $                   & $2.8 $                   & $0.63$                   & $3.2$                    \\
\mcol{2}{Charged kaon detection asymmetry}                & $0.022$                  & $0.030$                  & $0.0041$                 & $0.025$                  \\
\mcol{2}{Amplitudes for backgrounds}                      &                          &                          &                          &                          \\
& Combinatorial \D                                        & $3.5 $                   & $3.4 $                   & $4.7$                    & $6.4$                    \\
& Random $Dh$                                             & $0.10$                   & $0.16$                   & $0.066$                  & $0.16$                   \\
& $\PB_s$ partially reconstructed                         & $0.59$                   & $0.59$                   & $0.15$                   & $0.73$                   \\
\mcol{2}{$r_{\Bpm \rightarrow \D \pipm}$}                 & $1.8 $                   & $1.9 $                   & $1.6$                    & $1.1$                    \\
\mcol{2}{Efficiency over the phase space}                 & $5.7 $                   & $0.35$                   & $6.9$                    & $0.31$                   \\
\mcol{2}{\CP asymmetry fit bias}                          & $^{+5.7}_{-0}$           & $^{+5.1}_{-0}$           & $^{+0}_{-1.3}$           & $^{+2.6}_{-0}$           \\
\hline
\\[-10pt]
\mcol{2}{Total experiment or fit related}                 & $^{+9.6}_{-7.8}$         & $^{+9.0}_{-7.4}$         & $^{+9.1}_{-9.2}$         & $^{+9.6}_{-9.2}$         \\
\\[-10pt]
\hline
\mcol{2}{Total model related}                             & $1.0$                    & $3.0$                    & $4.6$                    & $8.4$                    \\
\hline
\end{tabular}
\label{tab:syst:summary}
\end{center}
\end{small}
\end{table}

The fractions of signal and background
are estimated with a fit to the \Bpm candidate invariant mass distributions.
To find the systematic uncertainties in \xypm arising from the uncertainties in these fractions,
the shapes and yields of the individual mass PDF contributions are modified and the fit repeated.
The largest changes in \xypm arise from modifications to the cross-feed and total combinatorial background components.
The uncertainties are therefore evaluated by repeating the \CP asymmetry fit with the cross-feed and total combinatorial background yields
independently varied by their statistical uncertainties.

The yield of combinatorial \D background is estimated using wrong-sign
candidates selected from data.
The systematic uncertainties arising from these estimates are found by repeating the \CP asymmetry fit to data with the yields
varied by the statistical uncertainties shown in Table~\ref{tab:SigWindowYields}.
Corresponding variations in the random $Dh$ background yield are made, so that the total combinatorial background yield,
obtained from the \Bpm invariant mass fit, is unchanged.

In the \Bpm invariant mass fit, a component PDF for partially reconstructed {\mbox{$\Bpm \rightarrow \D (\rightarrow \KS \pip \pim ) \mu^{\pm} \nu$}} background is not included.
The systematic uncertainty arising from this omission is found by repeating the \CP asymmetry fit to data with a contribution from this background.
The upper limits on the yields and the mass functions are found by applying muon identification requirements to the bachelor tracks of data candidates, 
and are kept constant in the fit.

In the \CP asymmetry fit, the background fractions obtained from the invariant mass
fit to \Bpm candidates are used for both \Bp and \Bm candidates.
This neglects any detection asymmetries for the charged bachelor tracks.
The \CP asymmetry fit is repeated with the central value of the charged kaon asymmetry, $(-1.2 \pm 0.2)\%$~\cite{KaonDetAsymm},
introduced for the signal and background components where the bachelor is expected to be a kaon.

In the \CP asymmetry fit, combinatorial \D background candidates are assumed to be distributed non-resonantly over the phase space of the \DtoKspipi\ decay.
The \CP asymmetry fit is repeated with the \D\ decay model changed to the sum of a phase-space distribution and a $\Kstarpm (892) $ resonance;
the fractions of the two components are fixed by a study of the Dalitz plot projections of data.

The \D\ decay model included in the \CP asymmetry fit for random $Dh$ background candidates is an incoherent sum of the two \DtoKspipi\ decay amplitudes
because it is equally likely for a $\Dz$ or $\Dzb$ meson to be present in an event.
The \CP asymmetry fit is repeated with the decay model changed to include the central value of the $\Dz-\Dzb$ production asymmetry of $(-1.0 \pm 0.3)\%$~\cite{Dprodasymm}.

The yield of \mbox{$\PB_s \rightarrow \D \Kstar$} partially reconstructed background candidates is very low in the signal invariant mass region,
but in the \CP asymmetry fit the candidates are assumed to be distributed in the same way as the suppressed component of signal \BpmtoDKpm\ over the \DtoKspipi\ decay phase space
and could therefore appear in particularly sensitive regions.
To estimate the systematic uncertainty arising from the assumed distribution, the \CP asymmetry fit to data is performed with the \D\ decay model for this
background changed to the favoured component of the signal \BpmtoDKpm\ decay model.

In order to allow the candidate detection, reconstruction and selection efficiency variation across the phase space of the \DtoKspipi\ decay 
to be found from \BpmtoDpipm\ data candidates,
the amplitudes from the suppressed decays $\Bm \rightarrow \Dzb \pim$ and $\Bp \rightarrow \Dz \pip$ are assumed to be negligible.
The systematic uncertainty arising from this assumption is estimated by repeating the \CP asymmetry fit to data with
an additional term in the signal \BpmtoDpipm\ and cross-feed \BpmtoDKpm\ decay models, representing the suppressed decay amplitudes.
The values of $r_{\Bpm \rightarrow \D \pipm}$, $\delta_{\Bpm \rightarrow \D \pipm }$ and $\gamma$ are fixed in the additional term;
various $r_{\Bpm \rightarrow \D \pipm}$ and $\delta_{\Bpm \rightarrow \D \pipm }$ values are assumed
($r_{\Bpm \rightarrow \D \pipm}= 0.01$, $0.015 $; $\delta_{\Bpm \rightarrow \D \pipm }= 0^\circ$, $90^\circ$, $180^\circ$, $270^\circ$, $315^\circ$), 
but in all cases $\gamma$ is set to $70 ^{\circ}$.

The efficiency variation across the \DtoKspipi\ decay phase space is parametrised in the \CP asymmetry fit by a second-order polynomial function in the variables $m^2_{+}$ and $m^2_{-}$.
To estimate the uncertainty arising from this, the \CP asymmetry fit to data is repeated with the efficiency parametrisation fixed and variations of the polynomial coefficients made.
A fit with a third-order polynomial function is also performed, with the efficiency parameters and \xypm allowed to vary.
The changes in the values of \xypm, compared to the nominal results, are taken as the systematic uncertainties
arising from the efficiency parametrisation.

The \CP asymmetry fit is verified using $1000$ data-sized simulated pseudo-experiments.
In each experiment the number and distribution of candidates is generated according to the fit result from data.
The obtained values of \xypm show a small bias when compared to the values used for the simulation;
these biases are included as systematic uncertainties.

To estimate the systematic uncertainty arising from the choice of amplitude model description of the \DtoKspipi\ decay,
\CP asymmetry fits with alternative model descriptions are performed on large samples of simulated decays.
For each alternative model, one element (for example, a resonance parameter) of the nominal model is altered.
One million \BpmtoDpipm\ and one million \BpmtoDKpm\ decays are simulated with the
model used for the nominal \CP asymmetry fit, and with the Cartesian parameters fixed to the fit result.
For the nominal model and each alternative model, a \CP asymmetry fit to the \BpmtoDpipm\
sample is performed with the coefficients of each resonance of the model allowed to vary.
Values for the Cartesian parameters \xypm are then obtained from a \CP asymmetry fit to the
\BpmtoDKpm\ sample, with the resonance coefficients fixed from the results of the fit to the \BpmtoDpipm\ sample.
The signed differences in the values of \xypm from the nominal results are taken as the systematic uncertainties,
with the relative signs between contributions indicating full correlation or anti-correlation.

In the alternative models considered, the following changes, labelled (a)-(u), have been applied, 
resulting in the uncertainties summarised in Table~\ref{tb.syst.modelsyst}: \vspace{-0.23cm}
\begin{itemize}\itemsep0pt \parskip0pt \parsep0pt \topsep0pt \partopsep0pt
  \item[$-$] $\pi\pi$ S-wave: The $F$-vector model is changed to use two other solutions of the $K$-matrix (from a total of three) determined from fits to
    scattering data~\cite{Anisovich:2002ij} (a), (b). The slowly varying part of the non-resonant term of the $P$-vector is removed (c).
  \item[$-$] $K  \pi$ S-wave: The generalised LASS parametrisation, used to describe the
    $\Kstar_0 (1430)$ resonance, is replaced by a relativistic Breit-Wigner propagator with parameters taken from Ref.~\cite{Aitala:2002kr} (d).
  \item[$-$] $\pi\pi$ P-wave: The Gounaris-Sakurai propagator is replaced by a relativistic Breit-Wigner propagator (e).
  \item[$-$] $K  \pi$ P-wave: The mass and width of the $\Kstar (1680)$ resonance are varied by their uncertainties from Ref.~\cite{Aston:1987ir} (f)$-$(i).
  \item[$-$] $\pi\pi$ D-wave: The mass and width of the $f_2(1270)$ resonance are varied by their uncertainties from Ref.~\cite{PDG2012} (j)$-$(m).
  \item[$-$] $K  \pi$ D-wave: The mass and width of the $\Kstar_2 (1430)$ resonance are varied by their uncertainties from Ref.~\cite{PDG2012} (n)$-$(q).
  \item[$-$] The radius of the Blatt-Weisskopf centrifugal barrier factors, $r_{{\rm BW}}$, is changed from $1.5\,\GeV^{-1}$ to $0.0\,\GeV^{-1}$ (r) and $3.0\,\GeV^{-1}$ (s).
  \item[$-$] Two further resonances, $\Kstar (1410)$ and $\Prho(1450)$, parametrised with relativistic Breit-Wigner propagators,
    are included in the model (t).
  \item[$-$] The Zemach formalism used for the angular distribution of the decay products is replaced by the helicity formalism (u).
\end{itemize}

The total covariance matrix is determined to be
\begin{equation}
V_{\mathrm{model}} =
\bordermatrix{
  &  x_-   &   y_-   &   x_+   &   y_+  \cr
  &  1.12  &  2.80   & - 0.95  & - 5.40 \cr
  &  2.80  &  8.89   & - 1.21  & -16.87 \cr
  & -0.95  & -1.21   &  21.59  &   5.97 \cr
  & -5.40  & -16.87  &   5.97  &  69.87 \cr
}\,\times\,10^{-6}
\end{equation}
resulting in total systematic uncertainties arising from the choice of amplitude model of
\begin{align*}
  \delta x_- &= 1.0 \times 10^{-3}, \\
  \delta y_- &= 3.0 \times 10^{-3}, \\
  \delta x_+ &= 4.6 \times 10^{-3}, \\
  \delta y_+ &= 8.4 \times 10^{-3}.
\end{align*}

\newcommand{\fProd} {$f_{\pi\pi,j}^{\mathrm{pr}}$\xspace}
\newcommand{\kus}[1]{\mrow{4}{$\Kstar(#1)$\xspace}}
\newcommand{\kts}[1]{\mrow{4}{$\Kstar_2(#1)$\xspace}}
\newcommand{\ftw}[1]{\mrow{4}{$f_2(#1)$\xspace}}
\newcommand{\kstl}  {$\Kstar(1410)$\xspace}
\newcommand{\rhol}  {$\Prho(1450)$\xspace}
\newcommand{\BW}    {{\rm BW}}
\newcolumntype{R}{>{$}r<{$}}

\newcommand{\GS} {Gounaris-Sakurai}
\newcommand{\RBW}{relativistic Breit-Wigner}
\newcommand{\gLASS}{Generalised LASS}

\newcommand{\tr}[1]{\mrow{2}{#1}}

\begin{table}[tb]
  \begin{small}
  \caption{\small Model related systematic uncertainties for each alternative model.  The relative signs
  indicate full correlation or anti-correlation.}
  \begin{center}
  \begin{tabular}{lllRRRR}
    \hline
             & \mcol{2}{Description}                 & \delta x_- \TMT & \delta y_- \TMT & \delta x_+ \TMT & \delta y_+ \TMT \\
    \hline
    (a)      & \mcol{2}{$K$-matrix 1st solution}     &       -0.1      &        0.04     &        0.3      &       -2        \\
    (b)      & \mcol{2}{$K$-matrix 2nd solution}     &       -0.09     &       -0.3      &        0.1      &       -0.5      \\
    \hline
    (c)      & \mcol{2}{Remove slowly varying}       &       -0.1      &       -0.3      &        0.1      &       -0.8      \\
             & \mcol{2}{part in $P$-vector}          &                 &                 &                 &                 \\
    \hline
    \tr{(d)} & \mcol{2}{\gLASS}                      &  \tr{$-0.7$}    &  \tr{$-2$}      &  \tr{$ 3$}      &  \tr{$ 7$}      \\
             & \mcol{2}{$\to$ \RBW}                  &                 &                 &                 &                 \\
    \tr{(e)} & \mcol{2}{\GS}                         &  \tr{$ 0.08$}   &  \tr{$-0.8$}    &  \tr{$ 0.1$}    &  \tr{$ 0.8$}    \\
             & \mcol{2}{$\to$ \RBW}                  &                 &                 &                 &                 \\
    \hline
    (f)      & \kus{1680} & $m + \delta m$           &       -0.06     &       -0.6      &        0.2      &        0.3      \\
    (g)      &            & $m - \delta m$           &       -0.1      &       -0.2      &       -0.1      &       -1        \\
    (h)      &            & $\Gamma + \delta \Gamma$ &       -0.06     &       -0.4      &       -0.05     &       -0.4      \\
    (i)      &            & $\Gamma - \delta \Gamma$ &       -0.2      &       -0.3      &        0.3      &       -0.5      \\
    \hline
    (j)      & \ftw{1270} & $m + \delta m$           &       -0.1      &       -0.3      &        0.1      &       -0.5      \\
    (k)      &            & $m - \delta m$           &       -0.1      &       -0.4      &        0.09     &       -0.5      \\
    (l)      &            & $\Gamma + \delta \Gamma$ &       -0.1      &       -0.3      &        0.08     &       -0.5      \\
    (m)      &            & $\Gamma - \delta \Gamma$ &       -0.1      &       -0.4      &        0.1      &       -0.5      \\
    \hline
    (n)      & \kts{1430} & $m + \delta m$           &       -0.08     &       -0.4      &        0.08     &       -0.4      \\
    (o)      &            & $m - \delta m$           &       -0.1      &       -0.3      &        0.1      &       -0.5      \\
    (p)      &            & $\Gamma + \delta \Gamma$ &       -0.1      &       -0.4      &        0.07     &       -0.4      \\
    (q)      &            & $\Gamma - \delta \Gamma$ &       -0.1      &       -0.3      &        0.1      &       -0.5      \\
    \hline
    (r)      & \mcol{2}{$r_{\BW} = 0.0 \GeV^{-1}$}   &       -0.2      &       -0.4      &       -0.1      &       -0.3      \\
    (s)      & \mcol{2}{$r_{\BW} = 3.0 \GeV^{-1}$}   &       -0.3      &       -0.3      &        1        &       -0.4      \\
    (t)      & \mcol{2}{Add \kstl and \rhol}         &       -0.1      &       -0.3      &        0.02     &       -0.7      \\
    (u)      & \mcol{2}{Helicity formalism}          &       -0.5      &       -2        &       -3        &        4        \\
    \hline
  \end{tabular}
  \end{center}
  \end{small}
  \label{tb.syst.modelsyst}
\end{table}

Table~\ref{tab:syst:summary} summarises the systematic uncertainties arising from all sources.
Except for the uncertainty due to the fit bias, the absolute values of the uncertainties 
are added in quadrature (assuming no correlation) to obtain the total experiment or fit related uncertainties.
The \CP asymmetry fit bias is considered as a one-sided uncertainty and is included in the quadrature sum on that side only.
The model related systematic uncertainty is also shown in the table, for comparison.

\section{Constraints on $\boldsymbol{\gamma}$, \textbf{\textit{r$_\textbf{\textit{B}}$}} and $\boldsymbol{\delta}_\textbf{\textit{B}}$}
\label{sec:interpretation}
The results for the \CP violation observables \xypm
 are used to place constraints on the values of $\gamma$, $r_{\B}$ and $\delta_{\B}$,
adopting the procedure described in Refs.~\cite{Aihara:2012aw, LHCb-PAPER-2012-027}.

\begin{figure}[h]
  \begin{center}
    \includegraphics[width=0.75\textwidth]{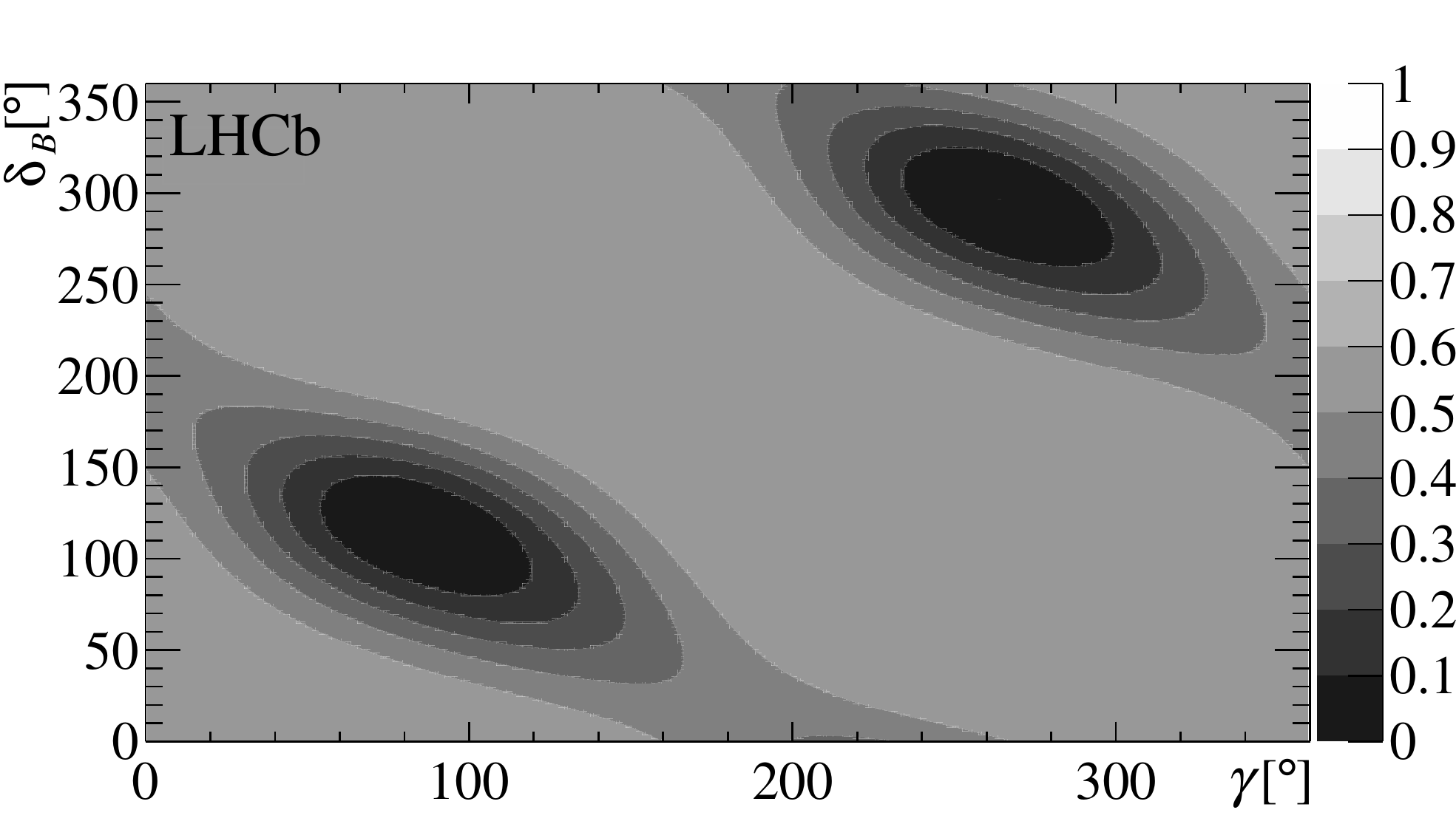}\\
    \includegraphics[width=0.75\textwidth]{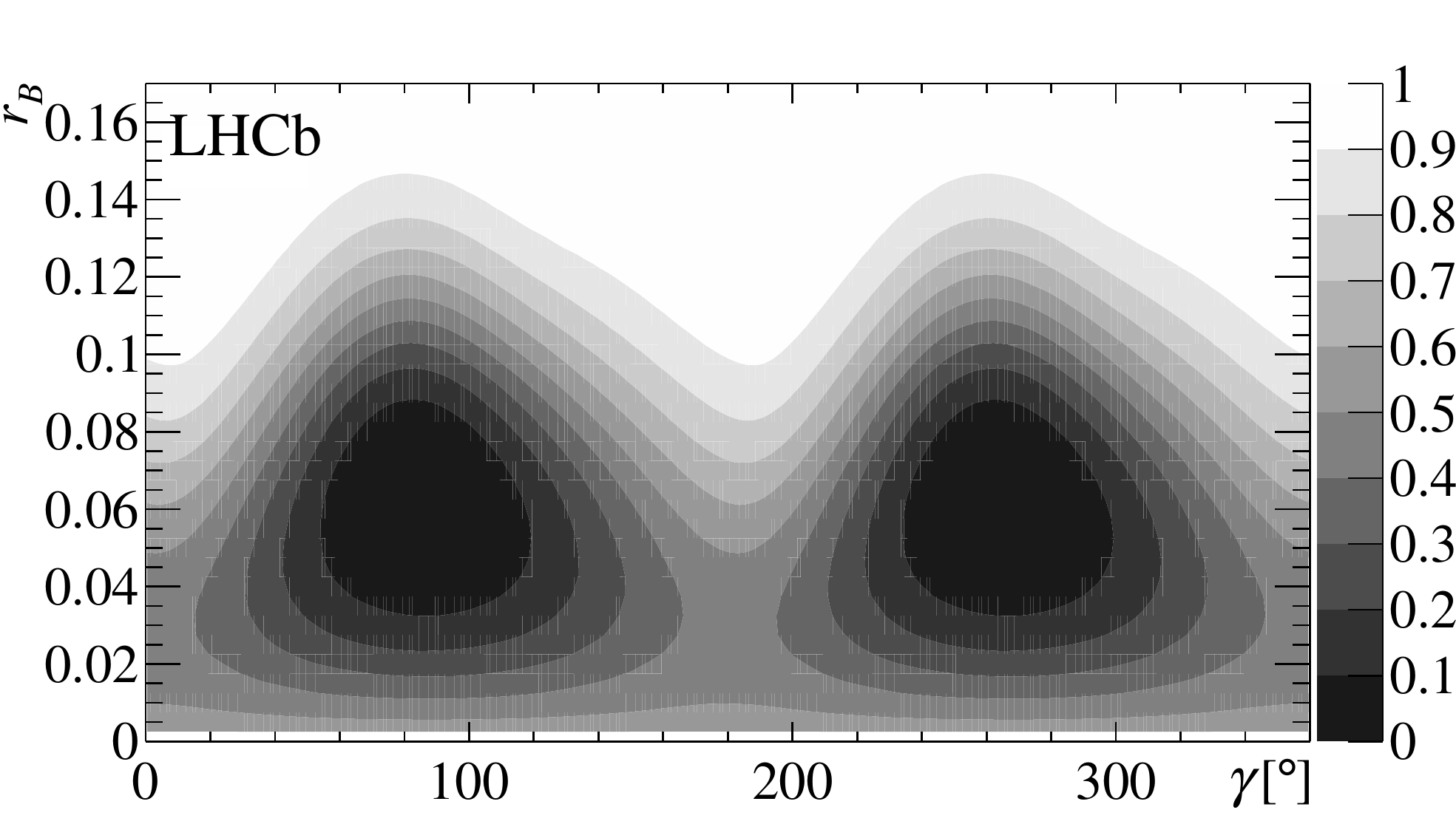}
    \caption{Projections of the p-value regions onto the $(\gamma, \delta_{\B})$ 
      and $(\gamma, r_{\B})$ planes with all sources of uncertainty taken into account.}
    \label{fg.inter.statsyst}
  \end{center}
\end{figure}

There is a two-fold ambiguity in the solution for $\gamma$, $r_{\B}$ and $\delta_{\B}$;
choosing the solution that satisfies $(0 < \gamma < 180)^\circ$ leads to the results
\begin{align*}
  \gamma &= (84^{+49}_{-42})^\circ,\\
  r_{\B} &= 0.06 \pm 0.04,\\
  \delta_{\B}&= (115^{+41}_{-51})^\circ,
\end{align*}
where the uncertainties include statistical, experimental systematic and model related systematic contributions.
Figure \ref{fg.inter.statsyst} shows the contours of p-value projected
onto the $(\gamma, \delta_{\B})$ and $(\gamma, r_{\B})$ planes.

\section{Effect of neutral \textbf{\textit{D}} meson mixing}
\label{sec:Dmix}
Assuming uniform lifetime acceptance, the measurements of the Cartesian parameters documented in this paper are corrected
for the effects of \D mixing as described in Ref.~\cite{Rama:2013voa},
\begin{align*}
x_{\pm}^{\text{corr}} &= x_{\pm} + \cfrac{y_{\text{mix}}}{2} ,\\
y_{\pm}^{\text{corr}} &= y_{\pm} + \cfrac{x_{\text{mix}}}{2} ,
\end{align*}
where $x_{\text{mix}}$ and $y_{\text{mix}}$ are the parameters of neutral \D meson mixing.

Since \CP violation in the charm sector has been neglected in the analysis, the world average values of the mixing parameters
without \CP violation ($x_{\text{mix}} = (0.53 ^{+0.16}_{-0.17}) \times 10^{-2}$, $y_{\text{mix}} = (0.67\pm0.09) \times 10^{-2}$)~\cite{ref:HFAG} 
are taken for correction, yielding the values
\begin{align*}
x_-^{\text{corr}} &=    +0.030 \pm 0.044  ~^{+0.010}_{-0.008}  \pm 0.001  \pm 0.00045, \\
y_-^{\text{corr}} &=    +0.016 \pm 0.048  ~^{+0.009}_{-0.007}  \pm 0.003  \pm 0.00085, \\
x_+^{\text{corr}} &=    -0.081 \pm 0.045  \pm 0.009           \pm 0.005  \pm 0.00045,  \\
y_+^{\text{corr}} &=    -0.029 \pm 0.048  ~^{+0.010}_{-0.009}  \pm 0.008  \pm 0.00085,
\end{align*}
where the first uncertainty is statistical, the second systematic, the third arises from the
\D decay amplitude model and the fourth is the uncertainty associated with the values of the mixing parameters.
The change in the value of $\gamma$ due to this correction is less than $1^\circ$.

\section{Conclusions}
\label{sec:Conclusions}
Candidate $\Bpm\rightarrow\D (\rightarrow \KS \pip \pim) \Kpm$ decays
are used to perform an amplitude analysis incorporating a model description of the \DtoKspipi\ decay.
The data used correspond to an integrated luminosity of $1\invfb$, 
recorded by \lhcb at a centre-of-mass energy of $7\tev$ in $2011$.

The resulting values of the \CP violation observables \mbox{$x_{\pm} = r_{\B} \cos{(\delta_{\B} \pm \gamma)}$} and
\mbox{$y_{\pm} = r_{\B} \sin{(\delta_{\B} \pm \gamma)}$} are
\begin{align}
x_- &=   +0.027  \pm 0.044 ~^{+0.010}_{-0.008} \pm 0.001, \notag \\
y_- &=   +0.013  \pm 0.048 ~^{+0.009}_{-0.007} \pm 0.003, \notag \\
x_+ &=   -0.084  \pm 0.045 \pm 0.009           \pm 0.005, \notag \\
y_+ &=   -0.032  \pm 0.048 ~^{+0.010}_{-0.009} \pm 0.008, \notag
\end{align}
where in each case the first uncertainty is statistical, the second systematic and the third 
is due to the choice of amplitude model used to describe the $\D \rightarrow \KS \pip \pim$ decay.
The results place constraints on the magnitude of the ratio of the
interfering \Bpm\ decay amplitudes,
the strong phase difference between them 
and the CKM angle $\gamma$, giving
the values $r_{\B} = 0.06 \pm 0.04$,
$\delta_{\B} = (115^{+41}_{-51})^\circ$ and 
$\gamma = (84^{+49}_{-42})^\circ$.
Neutral \D meson mixing has a negligible effect on the parameters $r_{\B}$, $\delta_{\B}$ and $\gamma$.

These results are consistent with, complementary to, and cannot be combined with, 
those obtained by the LHCb model-independent analysis of the same data set~\cite{LHCb-PAPER-2012-027}.
The results are also consistent with world average values~\cite{CKMfitter, UTfit}.

\section*{Acknowledgements}
\noindent We express our gratitude to our colleagues in the CERN
accelerator departments for the excellent performance of the LHC. We
thank the technical and administrative staff at the LHCb
institutes. We acknowledge support from CERN and from the national
agencies: CAPES, CNPq, FAPERJ and FINEP (Brazil); NSFC (China);
CNRS/IN2P3 (France); BMBF, DFG, HGF and MPG (Germany); SFI (Ireland); INFN (Italy); 
FOM and NWO (The Netherlands); MNiSW and NCN (Poland); MEN/IFA (Romania); 
MinES and FANO (Russia); MinECo (Spain); SNSF and SER (Switzerland); 
NASU (Ukraine); STFC (United Kingdom); NSF (USA).
The Tier1 computing centres are supported by IN2P3 (France), KIT and BMBF 
(Germany), INFN (Italy), NWO and SURF (The Netherlands), PIC (Spain), GridPP 
(United Kingdom).
We are indebted to the communities behind the multiple open 
source software packages on which we depend. We are also thankful for the 
computing resources and the access to software R\&D tools provided by Yandex LLC (Russia).
Individual groups or members have received support from 
EPLANET, Marie Sk\l{}odowska-Curie Actions and ERC (European Union), 
Conseil g\'{e}n\'{e}ral de Haute-Savoie, Labex ENIGMASS and OCEVU, 
R\'{e}gion Auvergne (France), RFBR (Russia), XuntaGal and GENCAT (Spain), Royal Society and Royal
Commission for the Exhibition of 1851 (United Kingdom).

\addcontentsline{toc}{section}{References}
\setboolean{inbibliography}{true}
\bibliographystyle{LHCb}
\bibliography{mybib}

\newpage
\centerline{\large\bf LHCb collaboration}
\begin{flushleft}
\small
R.~Aaij$^{41}$, 
B.~Adeva$^{37}$, 
M.~Adinolfi$^{46}$, 
A.~Affolder$^{52}$, 
Z.~Ajaltouni$^{5}$, 
J.~Albrecht$^{9}$, 
F.~Alessio$^{38}$, 
M.~Alexander$^{51}$, 
S.~Ali$^{41}$, 
G.~Alkhazov$^{30}$
% ...

{\footnotesize \it
% ...
$ ^{5}$Clermont Universit\'{e}, Universit\'{e} Blaise Pascal, CNRS/IN2P3, LPC, Clermont-Ferrand, France\\
% ...
$ ^{9}$Fakult\"{a}t Physik, Technische Universit\"{a}t Dortmund, Dortmund, Germany\\
% ...
$ ^{30}$Petersburg Nuclear Physics Institute (PNPI), Gatchina, Russia\\
% ...
$ ^{37}$Universidad de Santiago de Compostela, Santiago de Compostela, Spain\\
$ ^{38}$European Organization for Nuclear Research (CERN), Geneva, Switzerland\\
% ...
$ ^{41}$Nikhef National Institute for Subatomic Physics, Amsterdam, The Netherlands\\
% ...
$ ^{46}$H.H. Wills Physics Laboratory, University of Bristol, Bristol, United Kingdom\\
% ...
$ ^{51}$School of Physics and Astronomy, University of Glasgow, Glasgow, United Kingdom\\
$ ^{52}$Oliver Lodge Laboratory, University of Liverpool, Liverpool, United Kingdom\\
% ...
}
\end{flushleft}

\end{document}